\documentclass[epjc3]{svjour3}  
\pdfoutput = 1
\usepackage{graphicx}
\usepackage{fix-cm}
\usepackage{float}
\usepackage{afterpage}
\usepackage{epsfig,cite}
\usepackage{amssymb}
\usepackage{amsmath}
\usepackage{dsfont}
\usepackage{multirow}
\usepackage{url}
\usepackage{xcolor}
\usepackage{float}
\usepackage{afterpage}
\usepackage{booktabs}
\usepackage[right=2.5cm]{geometry}
\usepackage{tikz}
\usetikzlibrary{arrows}
\usepackage{tikz-3dplot}
\usepackage{enumitem}
\usepackage{amsfonts}
\usepackage{pifont}
\usepackage{epstopdf}
\usepackage{epsfig}
\usepackage[framed]{ntheorem}
\usepackage{framed}
\usepackage{makeidx}
\usepackage{simplewick}
\usepackage{hyperref}
\usepackage{placeins}
\usepackage{color}
\usepackage{fancyvrb}
\usepackage{subcaption} 
\usepackage{pythonhighlight}

\usepackage{tocloft}
\usepackage{listings}
\DefineVerbatimEnvironment{Highlighting}{Verbatim}{commandchars=\\\{\}}

\newcommand{\be}{\begin{equation}}
\newcommand{\ee}{\end{equation}}
\newcommand{\bea}{\begin{eqnarray}}
\newcommand{\eea}{\end{eqnarray}}
\newcommand{\bi}{\begin{itemize}}
\newcommand{\ei}{\end{itemize}}
\newcommand{\ben}{\begin{enumerate}}
\newcommand{\een}{\end{enumerate}}

\def\frac#1#2{{{#1}\over {#2}}}
\def\gsim{\mathrel{\rlap{\lower4pt\hbox{\hskip1pt$\sim$}}
    \raise1pt\hbox{$>$}}}         
\def\lsim{\mathrel{\rlap{\lower4pt\hbox{\hskip1pt$\sim$}}
    \raise1pt\hbox{$<$}}}         

\newcommand{\draft}[1]{}

\numberwithin{equation}{section}
\numberwithin{figure}{section}
\numberwithin{table}{section}

\usepackage{tabularx}
\newcolumntype{C}[1]{>{\centering\arraybackslash}p{#1}}

\renewcommand{\vec}[1]{\textbf{#1}}

\newcommand{\colibri}{\texttt{Colibri}}

\newcommand{\web}{\href{https://hep-pbsp.github.io/colibri/}{\url{https://hep-pbsp.github.io/colibri/}}}

\usepackage{neuralnetwork}
\usepackage{xstring}
\usepackage{xpatch}
\usepackage{array}

\definecolor{codegreen}{rgb}{0,0.6,0}
\definecolor{codegray}{rgb}{0.5,0.5,0.5}
\definecolor{codepurple}{rgb}{0.58,0,0.82}
\definecolor{backcolour}{rgb}{0.95,0.95,0.92}
\lstdefinestyle{mystyle}{
    backgroundcolor=\color{backcolour},   
    commentstyle=\color{codegreen},
    keywordstyle=\color{magenta},
    numberstyle=\tiny\color{codegray},
    stringstyle=\color{codepurple},
    basicstyle=\ttfamily\footnotesize,
    breakatwhitespace=false,         
    breaklines=true,                 
    captionpos=b,                    
    keepspaces=true,                 
    numbers=left,                    
    numbersep=5pt,                  
    showspaces=false,                
    showstringspaces=false,
    showtabs=false,                  
    tabsize=2
}

\makeatletter
\def\thickhline{%
             \noalign{\ifnum0 =`}\fi\hrule \@height \thickarrayrulewidth \futurelet
             \reserved@a\@xthickhline}
\def\@xthickhline{\ifx\reserved@a\thickhline
                \vskip\doublerulesep
                \vskip -\thickarrayrulewidth
                \fi
                \ifnum0 =`{\fi}}
\xpatchcmd{\linklayers}{\nn@lastnode}{\lastnode}{}{}
\xpatchcmd{\linklayers}{\nn@thisnode}{\thisnode}{}{}
\makeatother
\newlength{\thickarrayrulewidth}
 \makeatletter
 \renewcommand{\section}{%
  \@startsection{section}
    {0}
    {\z@}
    {-21dd plus-8pt minus-8pt}
    {6dd}
    {\normalsize\bfseries\boldmath}%
}
\renewcommand{\subsection}{%
  \@startsection{subsection}
    {1}
    {\z@}
    {-21dd plus-8pt minus-14pt}
    {6dd}
    {\normalsize\bfseries\boldmath}%
}
\renewcommand{\subsubsection}{%
  \@startsection{subsubsection}
    {2}
    {\z@}
    {-21dd plus-8pt minus-20pt}
    {6dd}
    {\normalsize\bfseries\boldmath}%
}
\makeatother

\usepackage{pythonhighlight}
\journalname{Eur. Phys. J. C}
\begin{document}
\title{\hspace{-0.9cm}\vspace{-0.5cm}\includegraphics[width=0.4\textwidth]{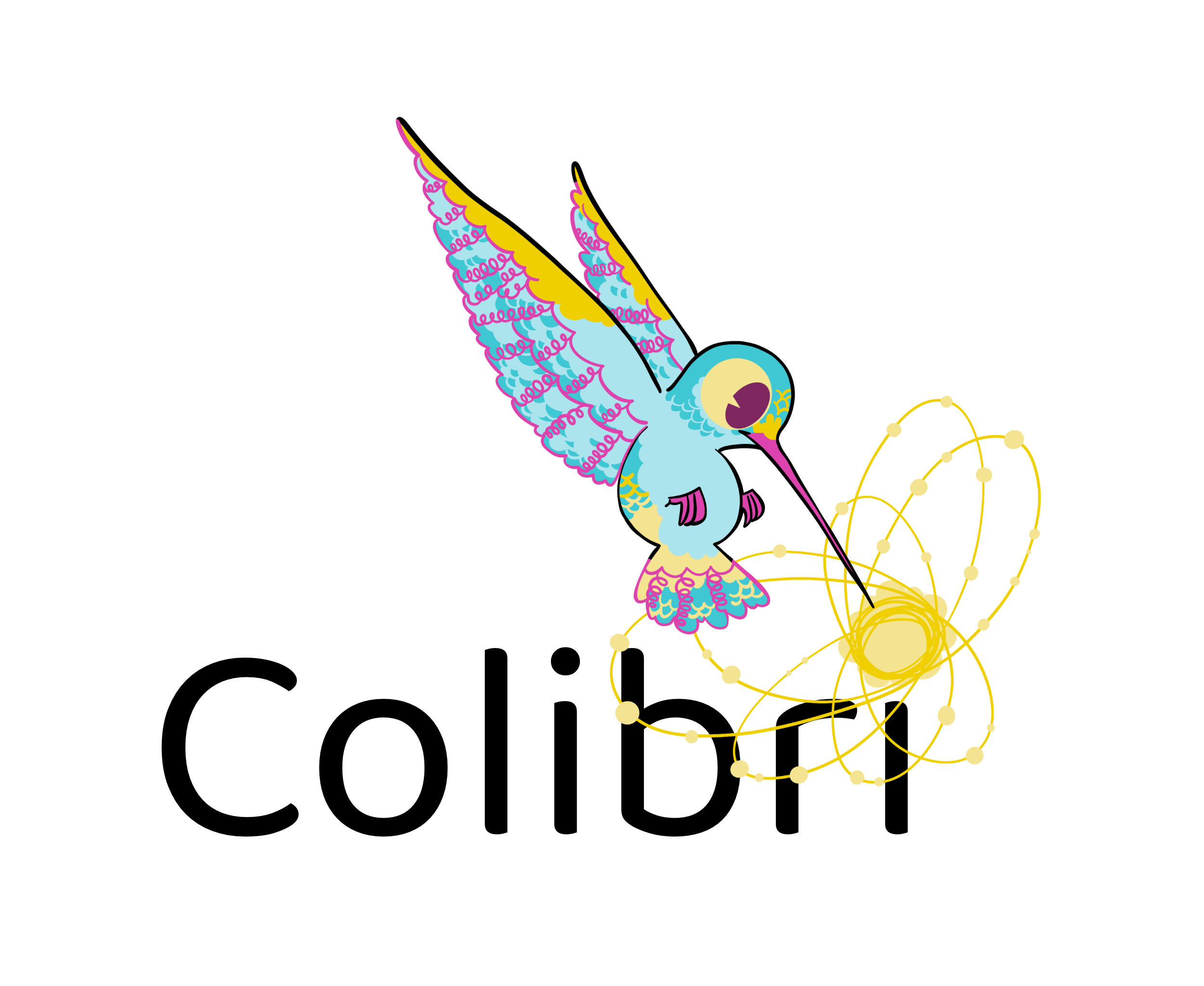}\\
A new tool for fast-flying PDF fits}
\author{Mark~N.~Costantini\thanksref{e0,addr1}
\and Luca~Mantani\thanksref{e1,addr2}\and 
James~M.~Moore\thanksref{e2,addr3}
\and Valentina~Sch\"utze~S\'{a}nchez \thanksref{e3,addr1}
\and Maria~Ubiali\thanksref{e4,addr1}}

\thankstext{e0}{e-mail: mnc33@cam.ac.uk}
\thankstext{e1}{e-mail: luca.mantani@uv.es}
\thankstext{e2}{e-mail: jmm232@cam.ac.uk}
\thankstext{e3}{e-mail: vcs32@cam.ac.uk}
\thankstext{e4}{e-mail: M.Ubiali@damtp.cam.ac.uk}

\setlength{\parindent}{0pt}

\institute{DAMTP, University of Cambridge, Wilberforce Road, Cambridge, CB3 0WA, UK \label{addr1} \and Lucy Cavendish College, Lady Margaret Road, Cambridge, CB3 0BU \label{addr3}\and Instituto de Fisica Corpuscular (IFIC), Universidad de Valencia-CSIC,
E-46980 Valencia, Spain\label{addr2}}

\date{Received: date / Accepted: date}

\maketitle

\setlength{\parindent}{15pt}

\begin{abstract}

We present \colibri{}, an open-source Python code that provides a general and flexible tool for PDF fits. 
The code is built so that users can implement their own PDF model, and use the built-in functionalities of \colibri{} for a fast computation of observables. It grants easy access to experimental data, several error propagation methodologies, including the Hessian method, the Monte Carlo replica method, and an efficient numerical Bayesian sampling algorithm. 
To demonstrate the capabilities of \colibri{}, we consider its simplest application: a polynomial PDF parametrisation. 
We perform closure tests using a full set of DIS data and compare the results of Hessian and Monte Carlo fits with those from a Bayesian fit.
We further discuss how the functionalities illustrated in this example can be extended to more complex PDF parametrisations. 
In particular, the Bayesian framework in \colibri{} provides a principled approach to model selection and model averaging, making it a valuable tool for benchmarking and combining different PDF parametrisations on solid statistical grounds.
\end{abstract}

\tableofcontents


\section{Introduction}
\label{sec:introduction}

A precise determination of the subnuclear structure of the proton is a central challenge in high-energy physics. 
Parton Distribution Functions (PDFs), which, to a first approximation, encode the momentum distributions of quarks, antiquarks and gluons inside the proton, 
are key ingredients in precision predictions for hadron collider experiments such as the Large Hadron Collider (LHC). 
Extracting PDFs from data is, however, a highly non-trivial statistical inverse problem that must reconcile theoretical 
constraints with a wealth of experimental measurements, together with their uncertainties and correlations~\cite{DelDebbio:2021whr,Amoroso:2022eow,Ubiali:2024pyg,Chiefa:2025loi}.  

Over the years, different inference strategies have been developed to address this task. 
The Hessian method, used in the first PDF fits producing error sets, 
provides an efficient framework for uncertainty estimate by approximating the likelihood surface 
as a quadratic function of parameters around the minimum, so that error propagation reduces to linear operations involving the covariance matrix of the fit parameters. 
This approach is computationally inexpensive for a moderately large parameter space and it is well suited for problems where the likelihood is close to Gaussian. 
However, it may produce unfaithful uncertainties when non-Gaussianities or parameter degeneracies are present. 
Several PDF fitting collaborations adopt the Hessian approach for error propagation~\cite{Hou:2019efy,Bailey:2020ooq,Alekhin:2017kpj,ATLAS:2021vod}, 
typically alongside a polynomial parametrisation of PDFs, and some collaborations introduce the concept of {\it tolerance}, which boils down to inflating 
uncertainties by a factor that accounts for inconsistencies among the various datasets\footnote{In the MSHT approach  
this is refined using a \textit{dynamical tolerance}, which is selected via a hyperoptimisation 
procedure~\cite{Martin:2009iq}.}~\cite{Pumplin:2009sc,Pumplin:2009bb,Harland-Lang:2024kvt,Barontini:2025lnl}. 

Monte Carlo replica methods, on the other hand, can be applied to an arbitrarily large parameter space, provided a cross-validation mechanism is in place. 
It relies on the generation of a large ensemble of {\it pseudodata replicas} that accurately reproduce the full experimental uncertainties and the correlations 
of the data used to determine PDFs~\footnote{In recent NNPDF studies the Monte Carlo replica sample incorporates theory uncertainties due to missing higher order 
uncertainties (MHOUs) and nuclear uncertainties into their PDF error propagation~\cite{NNPDF:2024dpb,Kassabov:2022orn,Ball:2021icz,NNPDF:2019vjt,NNPDF:2019ubu,Ball:2018twp}}. 
Each pseudo-data replica is fitted independently, and the statistical spread of the resulting PDF ensemble provides an estimate of 
the PDF uncertainties and of non-Gaussianities in the PDF ensemble. 
It was recently highlighted~\cite{Costantini:2024wby} that the Monte Carlo method faithfully estimates uncertainties in the linear regime, but it might be unreliable 
in the presence of non-linearities. The NNPDF~\cite{NNPDF:2021njg,NNPDF:2024djq,NNPDF:2024nan} 
and JAM~\cite{Moffat:2021dji} collaborations both use the Monte Carlo method to determine PDF uncertainties, the former alongside a redundant neural network parametrisation, 
the latter using a polynomial parametrisation for PDFs.

The Hessian and Monte Carlo approaches are both widely used in global PDF analyses, and there are well established ways to convert a Hessian PDF set into a Monte Carlo one and vice-versa~\cite{Watt:2012tq,Gao:2013bia,Carrazza:2015aoa,Carrazza:2015hva,Carrazza:2016htc,compressor}, 
that have been used to produce the two most recent PDF4LHC combinations~\cite{Butterworth:2015oua,PDF4LHCWorkingGroup:2022cjn} 
and the MSHT and NNPDF combination of the approximate N3LO sets~\cite{Cridge:2024icl}. 
By contrast, Bayesian inference, despite its ability to incorporate prior knowledge, quantify uncertainties 
probabilistically, and naturally accommodate theoretical constraints, has so far played only a limited role in practical 
PDF fitting. While some works have paved the way forward~\cite{Costantini:2025wxp,Candido:2024hjt,Capel:2024qkm,Albert:2024zsh,Aggarwal:2022cki,Gbedo:2017eyp}, a full-fledged PDF fit is yet to be produced, in part due to the absence of accessible, general-purpose tools.  

In this work we introduce \colibri{}, a new open-source platform written in Python, designed to perform global PDF fits with a unified treatment of inference. 
A key feature of our framework is its flexibility: \colibri{} can accommodate any PDF parametrisation defined by the user, and supports multiple inference methods, 
namely Hessian, Monte Carlo, and Bayesian fits, within 
a common infrastructure. This makes it possible to easily benchmark the same parametrisation across different methodologies, a capability that is unique to our approach and enhances both flexibility and robustness. Particularly novel is the integration of Bayesian inference, implemented through modern nested-sampling algorithms, which allow for the efficient exploration of 
high-dimensional parameter spaces and a probabilistic characterisation of uncertainties.  

While existing tools such as \texttt{xFitter}~\cite{xFitter:2022zjb,xFitterDevelopersTeam:2017xal} 
have played an important role in making PDF determinations more accessible, their scope is more restricted in key respects, as they 
only provide a limited set of parametrisations, and focus on two error-propagation strategies. 
The public release of the {\tt NNPDF} code~\cite{nnpdfcode}, on the other hand, has made the full-fledged NNPDF methodology, analysis tools, 
theory predictions and experimental data available to all users, making results reproducible and replicable. However, it is limited to the 
specific parametrisation and error propagation adopted by the NNPDF collaboration since the release of the {\tt NNPDF4.0} global fit~\cite{NNPDF:2021njg}.

In contrast, \colibri{}, builds on the availability of data and a fast interface with theory predictions from the {\tt NNPDF} public code~\cite{nnpdfcode,NNPDF:2021uiq} 
and it extends the current paradigm by providing a fully modular infrastructure, allowing users to freely choose both the parametrisation and the inference strategy. 
This comprehensive approach not only broadens the range of possible applications, but also enables direct 
and systematic comparisons between methodologies, a feature that is essential for advancing the robustness and reproducibility 
of global PDF analyses. Moreover, the Bayesian framework in \colibri{} enables the statistical combination of different PDF models 
based on their Bayesian evidence~\cite{Costantini:2025wxp}. This feature makes \colibri{} a valuable tool for benchmarking and combining PDF sets derived 
from diverse fitting methodologies on a rigorous statistical footing.

The paper is structured as follows. Section~\ref{sec:code} describes the architecture and implementation of \colibri{}, including 
the statistical algorithms and computational optimisations employed. In Section~\ref{sec:results}, we validate our framework 
with benchmark fits to existing datasets and compare the outcomes across inference methods. Finally, 
Section~\ref{sec:conclusions} summarises our findings and outlines future directions for development and applications.

\section{\colibri{}: a PDF building platform}
\label{sec:code}

In this section we describe the architecture and core functionalities of \colibri{}, highlighting how this Python framework allows users to implement arbitrary PDF parametrisations and fit them with a range of inference strategies.

\begin{figure}
    \centering
    \includegraphics[trim={0 3cm 0 3cm},clip, width=0.95\linewidth]{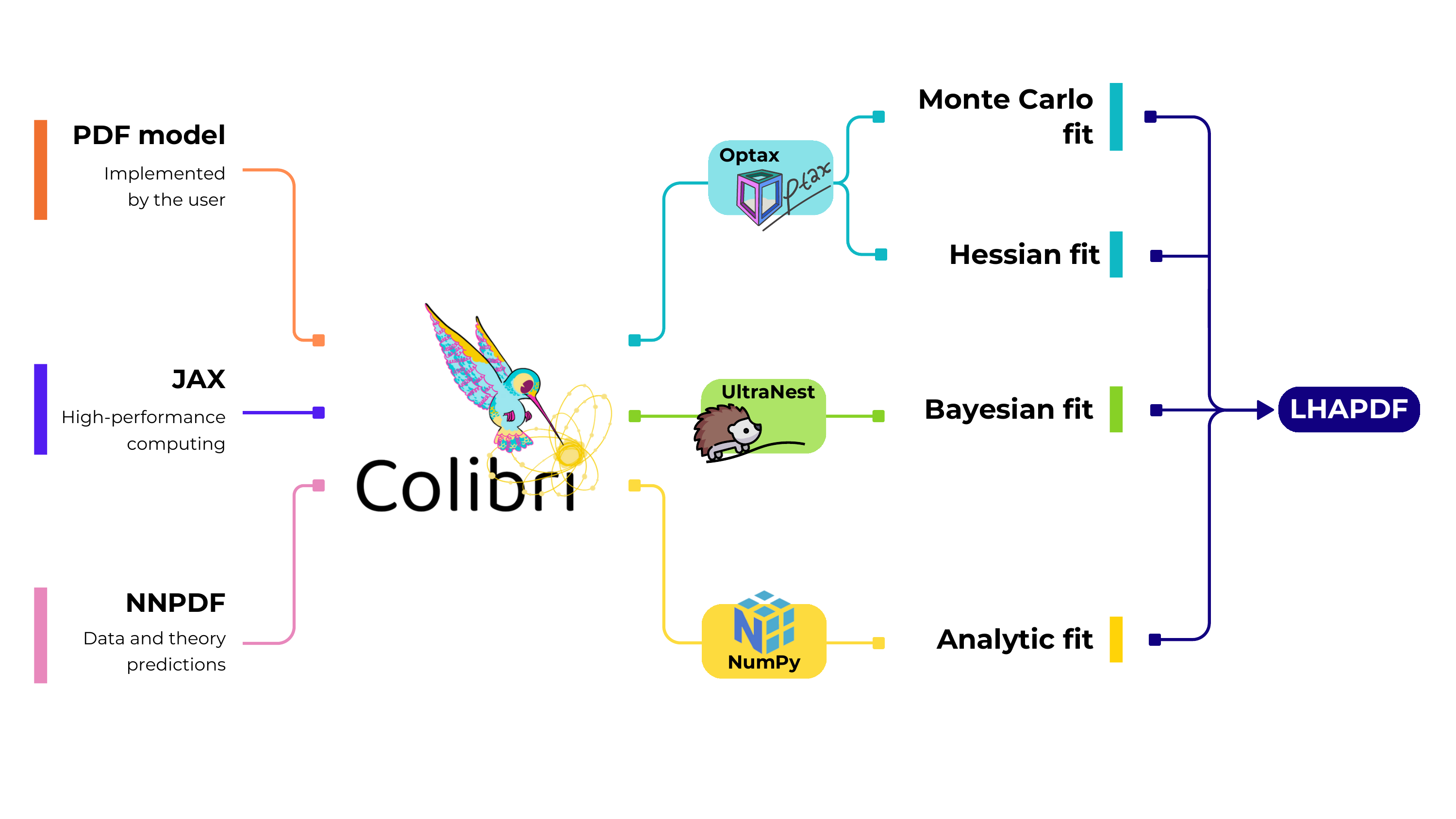}
    \caption{\colibri{}'s workflow: the code takes as input \textbf{(i)} a PDF model, which may be any arbitrary parametrisation implemented by the user, makes use of \textbf{(ii)} JAX~\cite{jax2018github}, 
    which provides high-performance array operations and native GPU support for fast
    computations, and inherits \textbf{(iii)} data and theory predictions from the {\tt NNPDF} public code~\cite{nnpdfcode,NNPDF:2021uiq}. 
    The code then performs a fit using a given inference method, which is specified by the user. At the time of release, the options are a Monte Carlo, Hessian, Bayesian or analytic fit. 
    In each case, the result is outputted in the LHAPDF format~\cite{Buckley:2014ana,LHAPDFurl}.}
    \label{fig:colibri-diagram}
\end{figure}

The design philosophy behind \colibri{} is built on three pillars: (i) {\it modularity}, as the tasks of defining a PDF model, 
constructing the likelihood, interfacing with data and theory predictions, and performing inference 
are separated into clear components with minimal assumptions about their specific form; 
(ii) {\it performance}, as the code is made to be fast and efficient by leveraging JAX's~\cite{jax2018github} high-performance array operations and native GPU support for fast computation; (iii) {\it universality}, as all models share the same inference methods as well as data and theory predictions. 
Modularity enables the user to benchmark different parametrisations under identical statistical conditions, 
and to test the impact of alternative inference strategies on the same model. High-level performance and universality enable users to perform reliable comparisons and rigorous methodological studies. Fig.~\ref{fig:colibri-diagram} shows a schematic overview of the code's workflow, showcasing the role that each of these modular building blocks play in its general functioning.

The section is organised as follows. 
We begin by introducing the \texttt{PDFModel} base class, which provides a uniform interface for all PDF parametrisations 
(Section~\ref{sec:pdf-model-class}). 
We then discuss the likelihood function and the implementation of theoretical and experimental constraints 
(Section~\ref{subsec:likelihood_function}). 
Next, we describe how data and theory predictions are incorporated, and how the forward map from PDF parameters to observables is built. 
Finally, we review the inference methods available in \colibri{}, including Hessian, Monte Carlo, and Bayesian error propagation.

\subsection{The \colibri{} PDF model class}
\label{sec:pdf-model-class}

To separate the definition of PDF parametrisations from their numerical inference, 
\colibri{} provides the abstract base class \texttt{PDFModel}. 
Listing~\ref{lst:pdfmodel} shows pseudo-code illustrating its structure. 
At a minimum, a \texttt{PDFModel} in \colibri{} must specify:
\begin{itemize}
  \item a list of model parameters, representing the degrees of freedom of the PDF;
  \item a method to map the model parameters to the values of the PDFs on a specified grid in momentum fraction $x$ for each parton flavour.
\end{itemize}

Such an abstraction allows users to implement a wide range of model architectures, 
from simple parametric forms to neural network based approaches, 
while leaving performance-critical tasks such as convolutions with pre-tabulated kernels 
and parameter sampling to optimised external engines. 

This means that, in practice, the user can implement a new parametrisation within \colibri{} by completing the abstract \texttt{PDFModel} class
and specifying the two required methods; parameter specification and forward map.
A detailed example of this procedure is given in the step-by-step tutorial outlined in~\ref{app:model}. We will now give an overview of what these methods are.

\begin{lstlisting}[language=Python,caption={Abstract PDFModel interface.},label={lst:pdfmodel}]
class PDFModel(ABC):
    """An abstract class describing the key features of a PDF model."""

    name = "Abstract PDFModel"

    @property
    @abstractmethod
    def param_names(self) -> list:
        pass

    @abstractmethod
    def grid_values_func(self, xgrid) -> Callable:
        """Produce a function that maps model parameters
        to PDF values on the grid `xgrid`.
        """
        pass

    def pred_and_pdf_func(
        self,
        xgrid,
        forward_map,
    ) -> Callable, Tuple:
        pdf_func = self.grid_values_func(xgrid)

        def pred_and_pdf(params, fast_kernel_arrays):
            pdf = pdf_func(params)
            predictions = forward_map(pdf, fast_kernel_arrays)
            return predictions, pdf

        return pred_and_pdf
\end{lstlisting}

\paragraph{Parameter specification}\mbox{}\\[0.3cm]
Every PDF model must declare the parameters to be fitted (e.g.\ normalisations, small-/large-$x$ exponents, polynomial coefficients, weights and biases of a neural network, 
weights of a linear combination, \ldots). 
These are listed in the \texttt{param\_names} property, which returns an ordered list of strings defining the parameter names in a fixed sequence.

\paragraph{Grid Evaluation Method}\mbox{}\\[0.3cm]
The core of the \texttt{PDFModel} class is the \texttt{grid\_values\_func} method,
which returns a JAX-compatible function~\cite{jax2018github},
\begin{align}
\mathbf{f} : \mathbb{R}^{N_{\theta}} \to \mathbb{R}^{N_{\rm fl} \times N_{\rm x}}, \qquad \vec{f} : \pmb{\theta} \mapsto \vec{f}(\pmb{\theta})
\end{align}
mapping an $N_{\theta}$-dimensional parameter vector $\boldsymbol{\theta}$ into the PDF values\footnote{Note that the $f_{\rm grid}$ method should actually return $x \text{PDF}$ (that is, the PDF multiplied by the momentum fraction $x$).} 
for each parton flavour index\footnote{These are $\gamma$, $\Sigma$, $g$, $V$, $V_3$, $V_8$, $V_{15}$, $V_{24}$, $V_{35}$, $T_3$, $T_8$, $T_{15}$, $T_{24}$ and $T_{35}$, since \colibri{} works in 
the evolution basis.} evaluated on the user-provided $x$-grid of length $N_x$.  In practice, for a standard PDF fit, the user only needs to define this method. 
The framework then automatically handles the construction of all the resources, such as the forward map from the parameters to the physical observables that enter the regression problem, 
i.e. the theory predictions, needed in a PDF fit.

\paragraph{Forward map and theory predictions}\mbox{}\\[0.3cm]
To compute physical observables (structure functions, cross sections, etc.), one must convolve the PDFs with partonic cross sections computed at a given perturbative order in the QCD and EW expansions. 
In \colibri{} this is handled via the \texttt{pred\_and\_pdf\_func} method, which takes again the $N_{\rm x}$-dimensional $x$-grid and a forward map that projects 
the PDF parameters to the space of physical observables. The method boils down to a function taking as input the PDF parameters and a tuple of fast-kernel (FK) matrices, 
and outputs the theoretical predictions, i.e. an explicit function $\mathbf{f}$ of the parameters $\pmb{\theta}$:
\begin{align}
(\boldsymbol{\theta},{\rm FK}) \mapsto \bigl(\mathrm{predictions}, \mathbf{f}(\boldsymbol{\theta})\bigr).
\end{align}
The matrix (FK) is called an FK-table in the NNPDF jargon, and provides a fast interpolation of the 
forward map, namely ${\rm (FK)}_{Ii} = \int c_I(x) \, p_i(x)\,dx$, where $c_I(x)$ are known functions computed in perturbation theory by convolving the PDF DGLAP evolution kernels and the partonic 
cross sections and $p_i(x)$ is an interpolation polynomial, relative to the point $x_i$.
Our notation reflects this convention. 
The FK arrays (i)~evaluate the PDF on the grid via \texttt{grid\_values\_func}, and (ii) feed the resulting $N_{\rm fl}\times N_{\rm x}$ array into 
the supplied \texttt{forward\_map} to yield a 1D vector of theory predictions for all data points.
Note that the prediction function is already implemented; however, the user is allowed to override it in its own PDF application 
if the specific model needs extra features.

%

\subsection{Likelihood function}
\label{subsec:likelihood_function}

Having defined a PDF model mapping $\mathbf{f}: \mathbb{R}^{N_{\theta}} \rightarrow \mathbb{R}^{N_{\rm fl}\times N_{x}},$ \colibri{} requires a likelihood function
\(\mathcal{L}(\mathbf{D}\mid\boldsymbol{\theta})\) that quantifies the agreement
between theory predictions and experimental data. 
The likelihood is the fundamental ingredient underlying all inference methods
implemented in our framework, namely Hessian, Monte Carlo, and Bayesian fits. Its specific form and implementation in \colibri{} are discussed in this section. Users may choose to employ the built-in implementation, with full control over its hyperparameters, or implement a custom version by overriding the \colibri{} likelihood function in their model whenever additional features are required.

\paragraph{Chi-Squared Likelihood}\mbox{}\\[0.3cm]
The basic form of the likelihood function 
is a chi-squared function:
\begin{equation}
  \label{eq:likelihood}
  \mathcal{L}(\mathbf{D}\mid\boldsymbol{\theta})
  = (\mathbf{D} - \mathrm{FK}[\mathbf{f}(\boldsymbol{\theta})])^{T}
    \;C_{t_0}^{-1}\;(\mathbf{D} - \mathrm{FK}[\mathbf{f}(\boldsymbol{\theta})]) \,,
\end{equation}
where \(\mathbf{D}\) is the data vector, \(\mathrm{FK}[\mathbf{f}(\boldsymbol{\theta})]\)
is the forward map, and \(C_{t_0}\) is the \(t_0\) covariance matrix used to avoid the d'Agostini bias when 
data have multiplicative uncertainties~\cite{Ball:2009qv}.
During a fit, it is possible to impose positivity and integrability constraints on PDFs as well as on observables. 

\paragraph{Positivity Constraints}\mbox{}\\[0.3cm]
Probability distributions for physical observables must necessarily be non-negative quantities. PDFs beyond LO, however, are not probabilities, and thus they may be negative. Now, it was recently shown in Refs.~\cite{Candido:2020yat,Candido:2023ujx} that, in the case of individual quark flavours and the gluon in the $\overline{\text{MS}}$ factorisation scheme, PDFs are indeed non-negative. We therefore allow users to impose this positivity condition along with the
constraint of positivity of physical cross sections discussed below. 

Positivity constraints on PDFs are implemented  similarly to NNPDF4.0~\cite{NNPDF:2021njg}, 
by adding extra Lagrange penalty terms to the likelihood function, namely
\begin{equation}
  \label{eq:pos_constraint}
  \mathcal{L}(\mathbf{D}\mid\boldsymbol{\theta})
  \;\longrightarrow\;
  \mathcal{L}(\mathbf{D}\mid\boldsymbol{\theta})
  \;+\;\Lambda_{\rm pos}\sum_{k}\sum_{i}
  \operatorname{Elu}_{\alpha}\bigl(-\tilde{f}_k(x_i,Q^{2})\bigr)\,,
\end{equation}
where by default \(Q^{2}=5\,\mathrm{GeV}^{2}\), and the points \(x_i\)
are ten values logarithmically spaced between \(5\times10^{-7}\) and
\(10^{-1}\), plus ten points linearly spaced between \(0.1\) and \(0.9\).
In addition, given that the positivity of $\overline{\text{MS}}$ PDFs is neither necessary nor sufficient in order to ensure cross section positivity, 
in order to exclude unphysical PDFs, we impose positivity of a number of cross sections, by means of Lagrange multipliers which penalise PDF
configurations leading to negative physical observables. Specifically, positivity can be imposed on the $F_2^u$, $F_2^d$, $F_2^s$ and $F_L$ 
structure functions and of the flavour-diagonal Drell-Yan rapidity distributions $\sigma_{\rm DY}^{u\bar{u}}$, $\sigma_{\rm DY}^{d\bar{d}}$ and $\sigma_{\rm DY}^{s\bar{s}}$.

\paragraph{Integrability Constraints}\mbox{}\\[0.3cm]
Integrability constraints are enforced similarly, by adding a term
\begin{equation}
  \label{eq:int_constraint}
  \mathcal{L}(\mathbf{D}\mid\boldsymbol{\theta})
  \;\longrightarrow\;
  \mathcal{L}(\mathbf{D}\mid\boldsymbol{\theta})
  \;+\;\Lambda_{\rm int}\sum_{k}\sum_{i}
  \bigl[x_i\,f_k(x_i,Q_{0}^{2})\bigr]^{2}\,,
\end{equation}
where \(Q_{0}\) is the parametrisation scale and the \(x_i\) run over the
small-\(x\) region of the FK-table grid (in practice, often only the
smallest \(x\) value is used to enforce this condition).

\subsection{Data and theory predictions}
\colibri{} provides a flexible platform that allows fitting PDF models to data that includes at least one incoming proton. 
The data is modelled in the framework of collinear QCD factorisation, where the scattering process is written as a convolution of the PDFs with perturbatively-computed, hard-scattering cross sections.
In this context, inferring the PDFs from experimental measurements is an inverse problem; the unknowns are the PDFs, and the forward model consists of the hard-scattering cross section combined with PDF evolution kernels,
commonly stored as FK-tables~\cite{Candido:2022tld} (fast-kernel-tables).
The data and FK-tables used in \colibri{} are inherited from the NNPDF framework~\cite{NNPDF:2021uiq}.

We distinguish two classes of forward maps based on whether the initial state involves one proton (Deep Inelastic
Scattering, DIS) or two protons (hadron-hadron collisions).

\paragraph{DIS data}\mbox{}\\[0.3cm]
DIS data is the most abundant data type in global PDF fits and is the most straightforward to model. For example, a measurement of the $F_{2}$ structure function, consisting of $N_{\rm dat}$ points, can be written as the contraction of two operators:
\begin{equation}
  \label{eq:dis_prediction}
  F_{2,i} \;=\; \sum_{j=1}^{N_{\rm fl}} \sum_{k=1}^{N_{x}} 
    \mathrm{FK}_{i,j,k}\;f_{j,k}\,,
\end{equation}
where the operator $\mathrm{FK}_{i,j,k}$ has shape $(N_{\rm dat},\,N_{\rm fl},\,N_{x})$, and $f_{j,k}$ is the $(N_{\rm fl}\times N_{x})$-dimensional grid representing the PDF values at the input scale $Q_{0}\,$.

\paragraph{Hadron-Hadron Predictions}\mbox{}\\[0.3cm]
Hadron-hadron collisions are more complicated to model than DIS data, as they involve the convolution of two incoming partons, each with their own PDF. An $N_{\rm dat}$-point measurement $\sigma$ of a hadron-hadron cross section can be written as
\begin{equation}
  \label{eq:had_prediction}
  \sigma_{i}
  = \sum_{j=1}^{N_{\rm fl}} \sum_{k=1}^{N_{\rm fl}}
    \sum_{l=1}^{N_{x}} \sum_{m=1}^{N_{x}}
    \mathrm{FK}_{i,j,k,l,m}\;f_{j,l}\;f_{k,m}\,,
\end{equation}
where the operator $\mathrm{FK}_{i,j,k,l,m}$ has shape
\((N_{\rm dat},\,N_{\rm fl},\,N_{\rm fl},\,N_{x},\,N_{x})\).

\subsection{Inference methods}
\label{subsec:inference}

In its release version, \colibri{} provides four inference strategies:
\begin{itemize}
  \item \textbf{Analytic fit}: computes the posterior mean and covariance of the parameters from the closed-form solution of a linear regression problem, yielding a Gaussian posterior.
  \item \textbf{Hessian method}: estimates parameter uncertainties from the curvature of the likelihood around the best-fit point, effectively linearising the problem in the vicinity of the minimum.
  \item \textbf{Monte Carlo replica method}~\cite{Costantini:2024wby}: constructs an ensemble of pseudodata replicas that reproduce experimental uncertainties, fits each independently, 
  and uses the statistical spread to approximate the posterior.
  \item \textbf{Bayesian inference}: explores the full posterior distribution of the parameters using modern nested-sampling algorithms, providing a principled probabilistic treatment of uncertainties.
\end{itemize}

The following subsections examine each method in more detail, highlighting their assumptions, strengths, and limitations.

\subsubsection{Analytic fits}
\label{subsec:analytic}

The analytic fit is the closed-form solution of a linear regression problem with Gaussian errors. 
It applies only when both the PDF model and the forward map are linear in the parameters, 
so that the likelihood is strictly quadratic. 
With a uniform prior on the parameters (and hence on the PDF values), 
the posterior is Gaussian, and its mean and covariance can be obtained directly from closed-form expressions, 
without any iterative optimisation or sampling.

This method cannot accommodate non-linear constraints such as positivity or integrability, 
and is therefore not suitable for realistic global PDF determinations. 
Nevertheless, it remains useful in several contexts:
\begin{itemize}
  \item \textbf{Fast benchmarks:} provides a lightweight means to validate new PDF parametrisations or data subsets 
  before running a full fit. 
  \item \textbf{Bayesian updating:} if a subset of data satisfies the linearity conditions, 
the resulting Gaussian posterior from that fit can serve as a prior in a subsequent fit 
to an uncorrelated dataset, thereby reducing the dimensionality of the sampling problem (see~\ref{app:bayesian_update} for more details).
  \item \textbf{Cross-checks:} analytic fits allow one to test the consistency of linear approximations against 
  more general inference strategies, highlighting the impact of non-linearities.
\end{itemize}

In practice, if the PDF model is linear in the parameters, the analytic fit is particularly useful for fitting linear DIS observables without constraints, 
where it yields closed-form posterior distributions at negligible computational cost. 
A detailed mathematical illustration of the method is provided in~\ref{app:bayes_lin_reg}.

\subsubsection{Hessian method}
\label{subsec:hessian}

The Hessian method is a widely used approach for parameter inference in PDF fits. 
It is based on a quadratic approximation of the likelihood around its minimum, 
so that uncertainties can be propagated through the covariance matrix of the fit parameters. 

In \colibri{}, the best-fit point is obtained by minimising the likelihood function using gradient-based optimisation algorithms provided by the \texttt{Optax} library~\cite{deepmind2020jax}. 
Once the minimum is found, the Hessian matrix $H$ of second derivatives with respect to the parameters is computed, and its inverse provides the covariance matrix $C$.
This covariance encodes the parameter uncertainties under the assumption of a locally Gaussian likelihood. 

As in traditional PDF analyses, it is possible to introduce a \emph{tolerance} factor $T>1$ to account for tensions or inconsistencies among datasets. 
In this case, the covariance matrix is rescaled as
\begin{equation}
  C \;\longrightarrow\; T^{2} C \,,
\end{equation}
effectively inflating the quoted parameter uncertainties~\cite{Pumplin:2001ct}. There are more modern ways to implement a dynamic tolerance, rather than a global one, 
that could be implemented in \colibri{}, see~\cite{Martin:2009iq,Lai:2010vv,Harland-Lang:2024kvt} for detailed discussions. 

For practical applications, Hessian PDF sets are distributed in terms of eigenvector directions of the covariance matrix. 
Diagonalising $C$ yields a set of eigenvalues and eigenvectors,
\begin{equation}
  C = V \Lambda V^{T},
\end{equation}
which define orthogonal directions in parameter space. 
The corresponding error sets are obtained by shifting the best-fit parameters along each eigenvector direction by $\pm \sqrt{\lambda_{i}}$, 
where $\lambda_{i}$ is the $i$-th eigenvalue. 
This representation is the standard format used in global PDF analyses, and is directly supported by \colibri{}.

The Hessian method is computationally efficient and provides a compact representation of PDF uncertainties. 
However, it relies on the quadratic approximation of the likelihood and may result in unfaithful uncertainties in the presence of non-linearities or parameter degeneracies. 
In \colibri{}, it serves both as a benchmark against which more general inference strategies can be compared, 
and as a practical tool for generating traditional Hessian PDF sets.

\subsubsection{Monte Carlo replica method}
\label{subsec:replica}

The Monte Carlo (MC) replica method is a widely used approach to estimate PDF uncertainties through repeated fits to pseudodata samples. 
In this approach, one generates $N_{\rm rep}$ replicas of the experimental dataset by sampling from a multivariate normal distribution with mean given by the central data values and covariance equal to the experimental covariance matrix. 
Each replica is then fitted independently, typically by minimising the chi-squared function using gradient-based optimisation algorithms, which in \colibri{} are provided by JAX~\cite{jax2018github} and Optax~\cite{deepmind2020jax}. 
The resulting ensemble of best-fit parameter sets approximately draws from the posterior distribution of the model parameters, and does so exactly in a linear regime. 
PDF uncertainties are then obtained from the statistical spread of this ensemble.

The MC replica method has several practical advantages:
\begin{itemize}
  \item \textbf{Robustness:} it makes minimal assumptions about the form of the likelihood surface, and does not require explicit computation of the Hessian matrix.
  \item \textbf{Flexibility:} it naturally incorporates correlations among data points and can accommodate non-linear models at the fitting stage.
  \item \textbf{Interpretability:} the replica ensemble provides a transparent representation of uncertainties that can be propagated to observables without additional approximations.
\end{itemize}

However, the method also has important limitations. 
As shown in Ref.~\cite{Costantini:2024wby}, the MC replica ensemble is formally equivalent to Bayesian posterior samples only in the case of linear models with Gaussian likelihoods. 
For non-linear parametrisations or forward maps, it can introduce biases and produce unreliable uncertainty estimates.
Moreover, the need to perform $N_{\rm rep}$ full fits makes the approach computationally more expensive than Hessian methods, especially for complex models.

For these reasons, \colibri{} implements the replica method primarily for benchmarking and for compatibility with existing PDF fitting practices, while Bayesian nested sampling (Section~\ref{subsec:bayesian}) is recommended as the principled approach for fully non-linear problems.

\subsubsection{Bayesian inference}
\label{subsec:bayesian}

Bayesian inference is the recommended strategy in \colibri{} for realistic PDF determinations. 
It provides a principled statistical foundation for uncertainty quantification, 
naturally incorporates prior knowledge, and enables robust model comparison. 
In this framework, the goal of a PDF fit is to characterise the full posterior distribution of the model parameters~$\boldsymbol{\theta}$,
\begin{equation}
  p(\boldsymbol{\theta} \mid \mathbf{D}) \;\propto\; \mathcal{L}(\mathbf{D}\mid\boldsymbol{\theta}) \,\pi(\boldsymbol{\theta}) \,,
\end{equation}
where $\mathcal{L}$ is the likelihood function and $\pi$ is the prior. 
The posterior encodes all information about the parameters given the data, 
allowing for a fully probabilistic treatment of uncertainties and correlations. 
Unlike the Hessian or MC replica methods, it does not rely on linear approximations or pseudo-data ensembles, 
and is therefore valid for arbitrary parametrisations and forward models.

A core strength of \colibri{} is the combination of user-defined PDF parametrisations with Bayesian inference. 
Through nested-sampling algorithms, this not only yields robust uncertainty quantification, 
but also provides the Bayesian evidence required for principled model selection and systematic comparisons of alternative parametrisations.

\paragraph{Implementation in \colibri{}}\mbox{}\\[0.3cm]
Sampling the posterior of high-dimensional PDF models is a challenging computational problem. 
\colibri{} addresses this using modern nested-sampling algorithms, as implemented in the \texttt{UltraNest} package~\cite{2021JOSS....6.3001B}. 
Nested sampling is well suited to PDF fits because it efficiently explores parameter spaces that may be multi-modal or strongly correlated, 
while also providing an estimate of the Bayesian evidence for principled model comparison. 
This makes Bayesian inference not only the most robust option for uncertainty quantification, but also a powerful tool for testing alternative PDF parametrisations within the same framework. 
The potential of this approach has already been demonstrated in our recent work~\cite{Costantini:2025wxp}, 
where dimensionality reduction was applied to the NNPDF neural network parametrisation. 
That study delivered the first realistic Bayesian PDF fit using \colibri{} and showed that model selection can be performed with minimal parametrisation complexity while maintaining excellent agreement with data.

\paragraph{Prior distributions}\mbox{}\\[0.3cm]
Bayesian inference requires the specification of a prior $\pi(\boldsymbol{\theta})$, 
which encodes information about the parameters before the data are taken into account. 
\colibri{}, at the time of the release, supports two built-in options:
\begin{itemize}
  \item \textbf{Uniform priors}, with configurable bounds for each parameter.
  \item \textbf{Gaussian priors}, defined by a mean vector and covariance matrix taken from the posterior of a previous fit.
\end{itemize}
The Gaussian option enables a Bayesian update (posterior--factorisation): 
when an earlier fit yields an approximately Gaussian posterior and the datasets are uncorrelated, 
that posterior can be re-used as the prior for a subsequent fit. 
\ref{app:bayesian_update} discusses the theoretical basis and domain of validity of this approach. 

In addition, users may implement fully customised priors by overriding the prior function in their model definition. 
This flexibility is essential for incorporating external information or theoretical constraints, and exemplifies the modularity of the \colibri{} framework. See Code Listing~\ref{lst:prior-overriding} for an example of how to override the \colibri{} built-in prior to specify a unit gaussian prior for the parameters.

\begin{lstlisting}[language=Python,caption={Example of how to override the prior built-in \colibri{}, specifying a unit gaussian prior for the parameters instead.},label={lst:prior-overriding}]
import jax
from colibri import bayes_prior
from colibri.utils import cast_to_numpy

def bayesian_prior(prior_settings, pdf_model):
    """
    Override of bayesian_prior:
    - if prior_distribution == "unit_gaussian", returns a unit Gaussian prior
    - otherwise falls back to colibri.bayes_prior
    """

    if prior_settings.prior_distribution == "unit_gaussian":

        @cast_to_numpy
        @jax.jit
        def prior_transform(cube):
            # transform [0,1] -> N(0,1)
            return jax.scipy.stats.norm.ppf(cube)

        return prior_transform

    # fallback to the normal bayesian_prior
    return bayes_prior.bayesian_prior(prior_settings, pdf_model)
\end{lstlisting}
\section{Case study: a simple \colibri{} fit with Les Houches PDFs}
\label{sec:results}
\begin{figure}
    \centering
    \begin{subfigure}[t]{0.48\linewidth}
        \centering
        \includegraphics[clip,width=\linewidth]{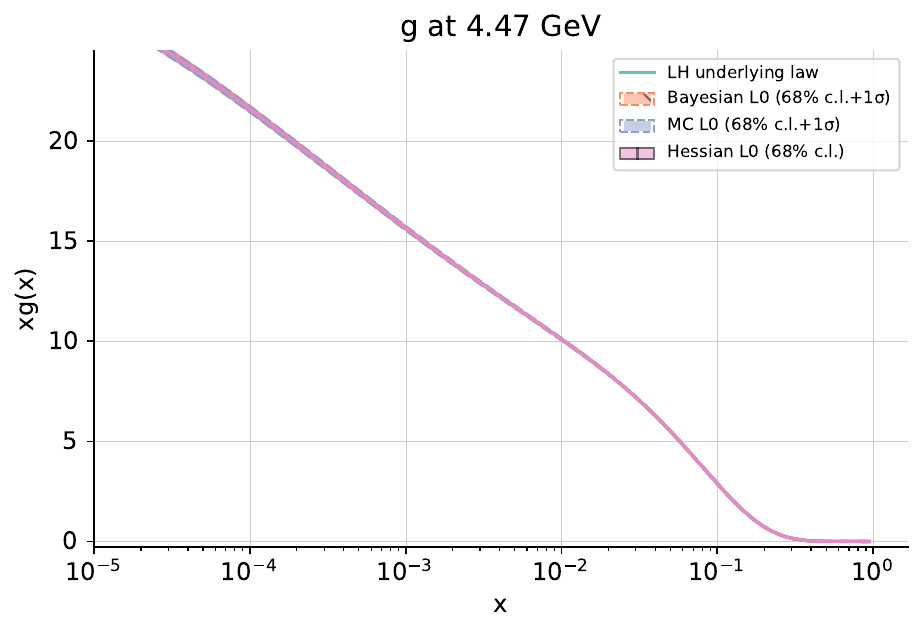}
    \end{subfigure}
    \hfill
    \begin{subfigure}[t]{0.48\linewidth}
        \centering
        \includegraphics[clip,width=\linewidth]{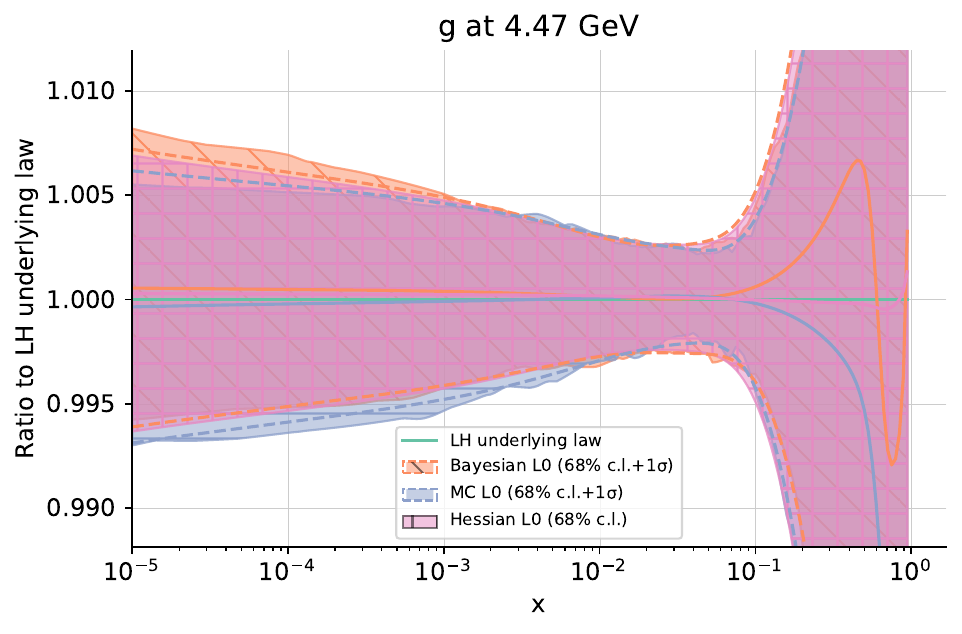}
    \end{subfigure}
    \caption{A gluon PDF fit resulting from a level~0 closure test computed with the Monte Carlo replica method (blue), Bayesian inference (orange), and the Hessian method (red). 
The green line shows the underlying law to be recovered, which in this case is the Les Houches parametrisation with best-fit parameter values from Ref.~\cite{Alekhin:2005xgg}. 
The right-hand panel shows the ratio to the underlying law.}
    \label{fig:gluon-level0-comparison}
\end{figure}

\begin{figure}
    \centering
    \begin{subfigure}[t]{0.48\linewidth}
        \centering
        \includegraphics[clip,width=\linewidth]{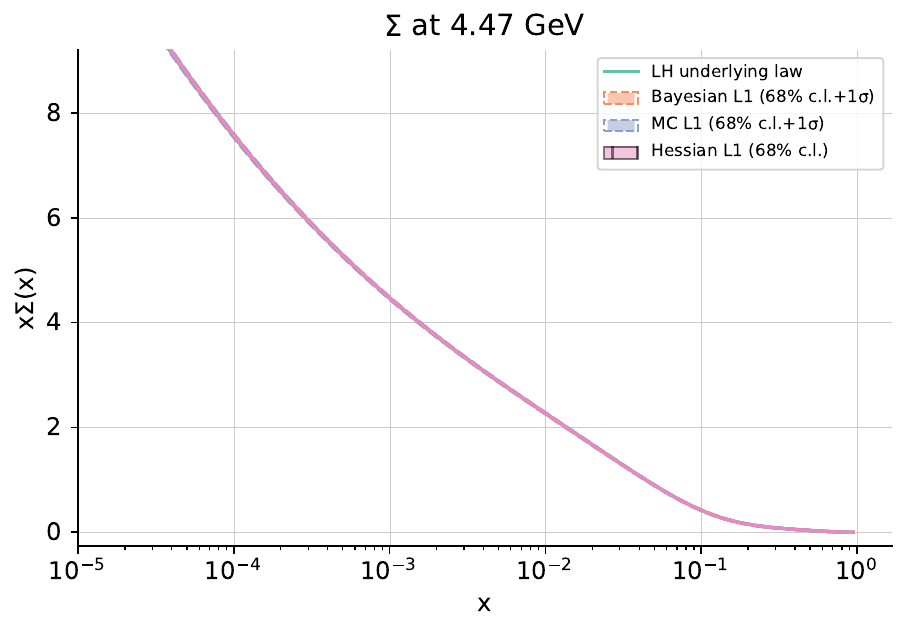}
    \end{subfigure}
    \hfill
    \begin{subfigure}[t]{0.48\linewidth}
        \centering
        \includegraphics[clip,width=\linewidth]{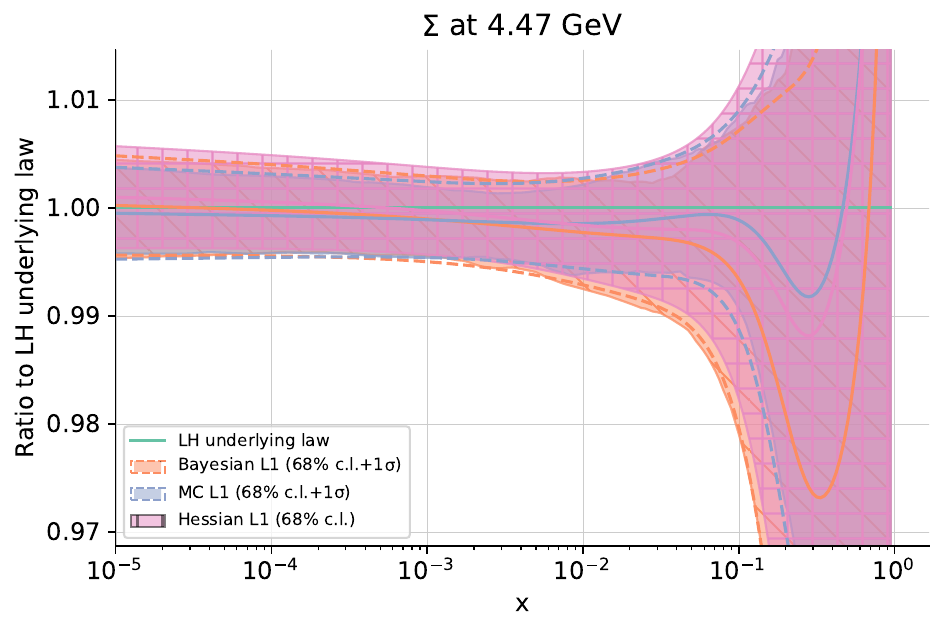}
    \end{subfigure}
    \caption{A $\Sigma$ PDF fit resulting from a level~1 closure test computed with the Monte Carlo replica method (blue), Bayesian inference (orange), and the Hessian method (pink). 
The green line shows the underlying law to be recovered, which in this case is the Les Houches parametrisation with best-fit parameter values from Ref.~\cite{Alekhin:2005xgg}. 
The right-hand panel shows the ratio to the underlying law.}
\label{fig:sigma-level1-comparison}
\end{figure}

To demonstrate the capabilities of \colibri{}, we present a set of benchmark fits performed with a simple model 
that was put forward in one of the Les Houches benchmarks to compare various PDF fitting methodologies~\cite{Alekhin:2005xgg}. 
The Les Houches model provides a simple polynomial parametrisation of PDFs under some assumptions that are explicitly spelled out in~\ref{app:les-houches-param}. 
In the evolution basis, the four independent PDFs are parametrised as:
\begin{align}
\label{eq:evolution-basis-set}
xf_g(x,Q_0) &= A_g\,x^{\alpha_g}\,(1 - x)^{\beta_g} \notag \\
xf_\Sigma(x,Q_0) &=  A_\Sigma\,x^{\alpha_\Sigma}\,(1 - x)^{\beta_\Sigma} \notag \\
x f_V &= x f_{u_v} + x f_{d_v} \\
&= A_{u_v}\,x^{\alpha_{u_v}}\,(1 - x)^{\beta_{u_v}}\, (1+\epsilon_{u_v}\sqrt{x}+\gamma_{u_v} x) + A_{d_v}\,x^{\alpha_{d_v}}\,(1 - x)^{\beta_{d_v}}(1+\epsilon_{d_v}\sqrt{x}+\gamma_{d_v} x) \notag \\
x f_{V_3} &= x f_{u_v} - x f_{d_v} \notag \\
&= A_{u_v}\,x^{\alpha_{u_v}}\,(1 - x)^{\beta_{u_v}}\, (1+\epsilon_{u_v}\sqrt{x}+\gamma_{u_v} x) - A_{d_v}\,x^{\alpha_{d_v}}\,(1 - x)^{\beta_{d_v}}(1+\epsilon_{d_v}\sqrt{x}+\gamma_{d_v} x).\notag
\end{align}
After applying the Les Houches parametrisation assumptions and sum rules (spelled out explicitly in~\ref{app:les-houches-param}, along with a detailed discussion of the rotation to the evolution basis and the expressions for the normalisation factor) we are left with 13 free parameters, namely $\alpha_g,\allowbreak\ \beta_g,\allowbreak\ \alpha_{u_v},\allowbreak\ \beta_{u_v},\allowbreak\ \epsilon_{u_v},\allowbreak\ \gamma_{u_v},\allowbreak\ \alpha_{d_v},\allowbreak\ \beta_{d_v},\allowbreak\ \epsilon_{d_v},\allowbreak\ \gamma_{d_v},\allowbreak\ \alpha_\Sigma,\allowbreak\ \beta_\Sigma$ and $A_g(A_\Sigma)$.
 
To showcase the performance of the new tool, we consider a fit to synthetic data, as performed in Refs.~\cite{NNPDF:2014otw,DelDebbio:2021whr}.
There are two such levels of data, namely
\begin{equation}
    {\bf D}^{L_0} = {\bf T[f(\pmb{\theta}^*)]},
\end{equation}
where $\pmb{\theta}^*$ are the ``true" PDF parameters, taken from some underlying law (in this case the best fit parameters determined in \cite{Alekhin:2005xgg}) 
that are used to generate the theory predictions for observables. This is called Level-0 data, and is nothing but the underlying law itself, built by
convolving the ``true" PDFs with partonic cross sections computed at a given perturbative order. Adding Gaussian noise generated from the
covariance matrix of the input data, we obtain Level-1 data, namely;
\begin{equation}
    {\bf D}^{L_1} = {\bf T[f(\pmb{\theta}^*)]} + {\bf \eta},
\end{equation}
where ${\bf \eta}\sim {\cal N}(0,C)$, and $C$ is the covariance matrix used in the fit. We fit it only to the set of DIS
data included in  the NNPDF4.0 analysis~\cite{NNPDF:2021njg}.
In a Level-0 test, synthetic data are generated directly from the underlying parametrisation without statistical fluctuations, 
so the goal of the fit is to recover the exact law. 
In a Level-1 test, synthetic data include statistical noise consistent with the experimental covariance, 
making the exercise closer to a realistic fit.

Figures~\ref{fig:gluon-level0-comparison} and~\ref{fig:sigma-level1-comparison} show representative results for the gluon and $\Sigma$ PDFs. 
In all cases, the underlying law (green) is well reproduced by Bayesian (orange), Monte Carlo replica (blue), and Hessian (pink) fits. 
As expected, the Level-0 closure test demonstrates near-perfect agreement with the generating function, 
while in the Level-1 test the fitted distributions track the law within the quoted uncertainties. 
The right-hand panels display the ratio to the ``truth", highlighting the consistency of the three approaches across the full $x$ range. 
In this simplified scenario the three methodologies yield comparable results, 
likely because the Gaussian approximation holds well. 
In more general settings, however, this need not be the case, 
and deviations between inference methods may occur.

\begin{table}[h]
\centering
\begin{tabular}{c|c|c|c}
    & Bayesian & MC & Hessian\\ \hline
Level 0 &  5.21$\times 10^{-4}$  &  1.76$\times 10^{-5}$  & 2.39$\times 10^{-5}$\\ \hline
Level 1 &    1.00   &   1.01  & 1.01
\end{tabular}
\caption{Comparison of Bayesian, Monte Carlo and Hessian $\chi^2$ values for Level 0 and Level 1 closure tests, where the underlying law was the Les Houches parametrisation model.}
\label{tab:bayes-mc}
\end{table}
Table~\ref{tab:bayes-mc} compares the $\chi^2$ values obtained with the Bayesian, Monte Carlo, and Hessian methods. 
All three strategies give compatible results in both closure tests, 
confirming the internal consistency of the framework.

\begin{figure}
    \centering
    \begin{subfigure}[t]{0.48\linewidth}
        \centering
        \includegraphics[clip,width=\linewidth]{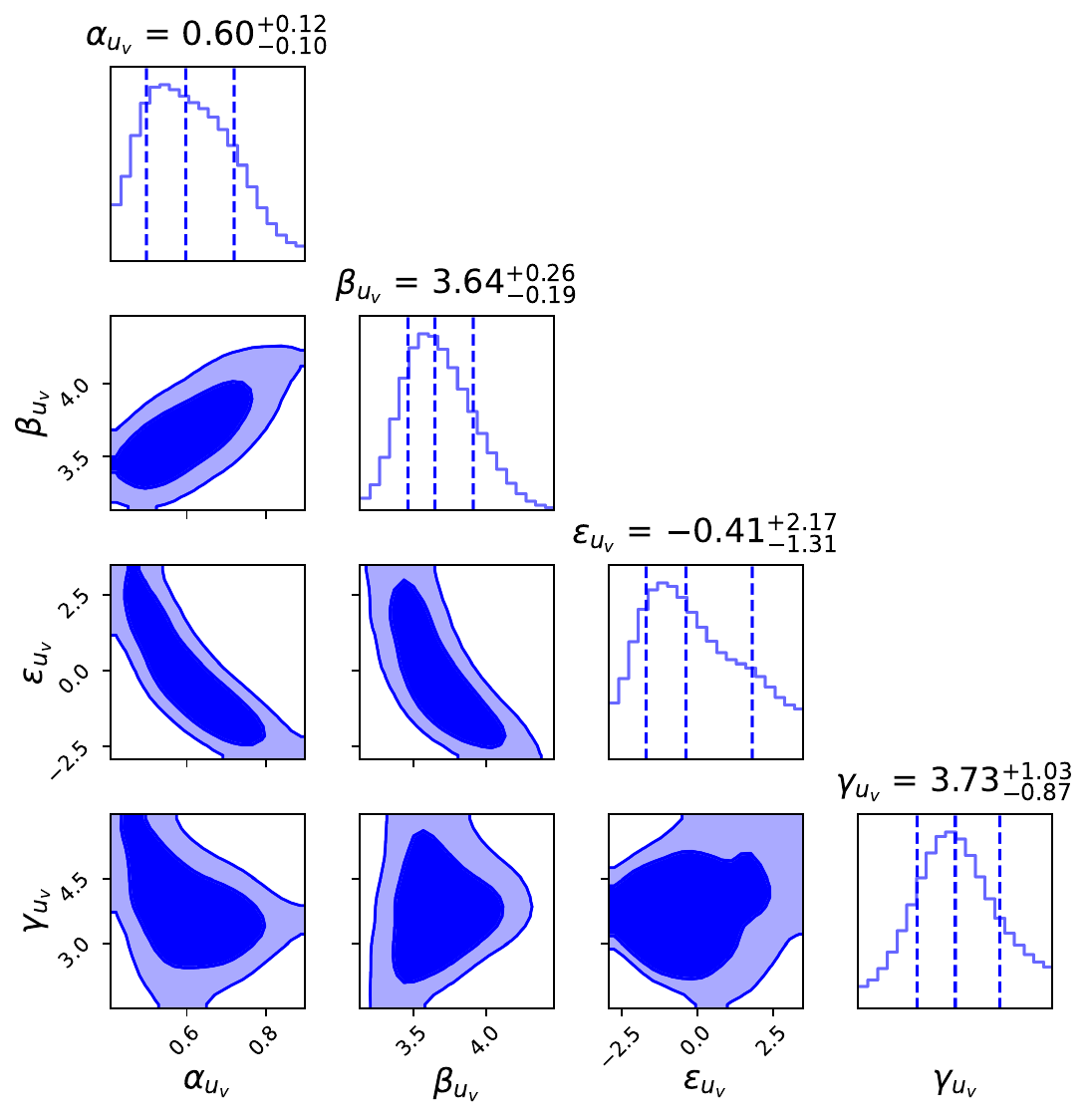}
    \end{subfigure}
    \hfill
    \begin{subfigure}[t]{0.48\linewidth}
        \centering
        \includegraphics[clip,width=\linewidth]{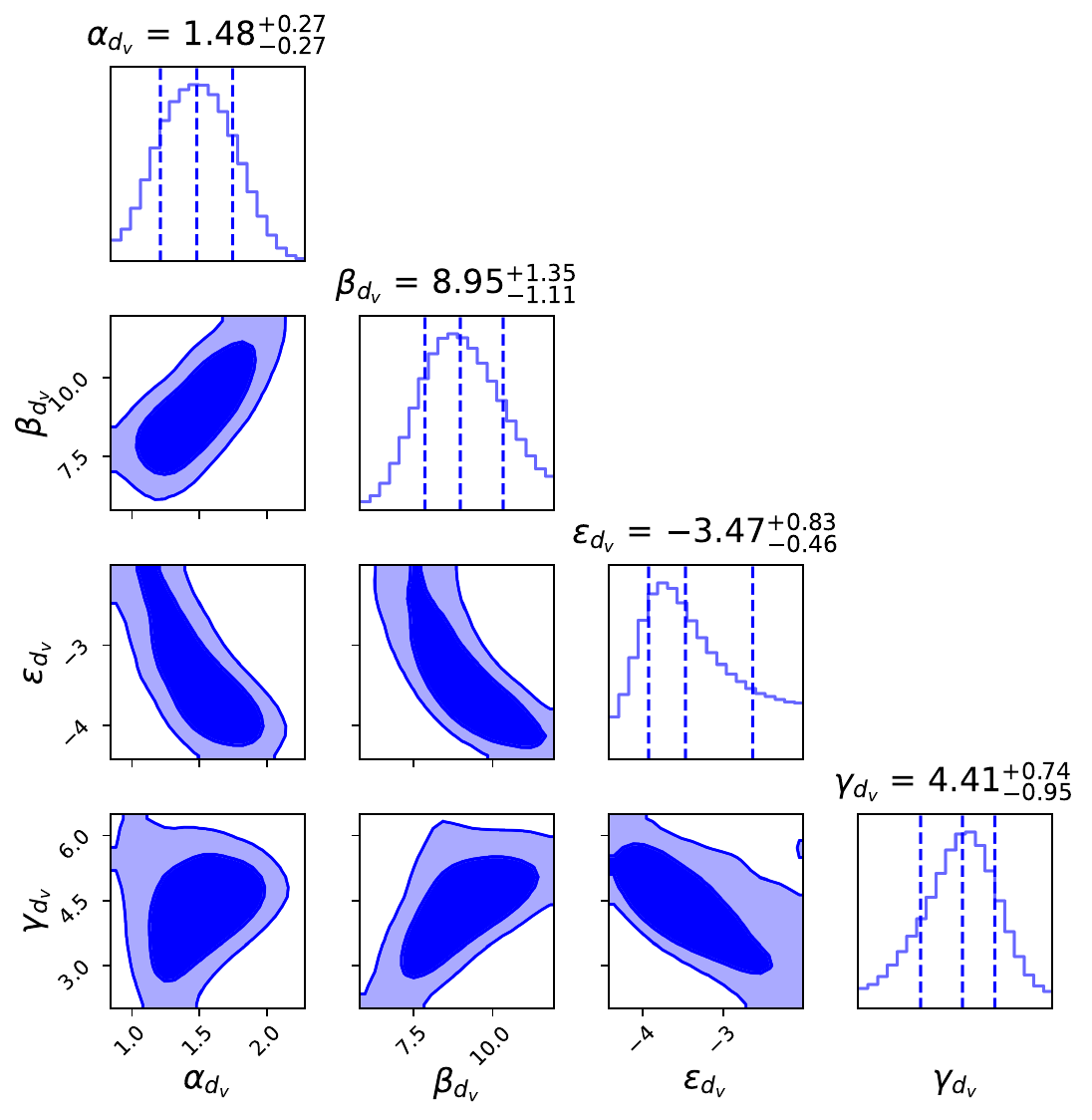}
    \end{subfigure}
    \caption{Example corner plots of posterior samples. 
The left panel shows the parameters associated with the $u$ valence quark, 
and the right panel those of the $d$ valence quark. 
In principle, analogous plots could be produced for all 13 parameters, 
but here we restrict to subsets for clarity and illustration.}
    \label{fig:corner-plot}
\end{figure}
An additional advantage of the Bayesian approach is that it yields direct samples from the posterior distribution of the PDF parameters. 
Beyond the reduced set of replicas exported in LHAPDF format, \colibri{} retains the full collection of posterior samples generated during the nested-sampling run. 
These samples can be analysed further to extract detailed information about the parameter space, for instance by producing corner plots that expose correlations and degeneracies among parameters. 
Figure~\ref{fig:corner-plot} shows an example corner plot from the Bayesian Level-1 closure test, illustrating how \colibri{} facilitates a transparent exploration of the multidimensional posterior beyond the one-dimensional PDF projections.

These benchmarks illustrate how \colibri{} makes it straightforward to perform and compare fits with different inference strategies within a single infrastructure. 
By applying them to the same dataset and parametrisation, 
one can systematically study the assumptions and limitations of each method and validate the robustness of PDF determinations.

\section{Conclusions}
\label{sec:conclusions}

In this work we have presented \colibri{}, a new open-source platform for parton distribution function determinations. 
The framework is designed around two central principles: user-defined PDF parametrisations and a unified treatment of inference. 
This modularity allows for direct comparisons between inference strategies, Hessian, Monte Carlo replicas, and Bayesian nested sampling, under identical conditions, thereby exposing their respective strengths and limitations. 
As illustrated in \ref{app:model}, we have shown how a PDF model can be implemented in \colibri{} by using the Les Houches parametrisation as a worked example. 
With this model, we performed closure tests that showcase the results of the three inference strategies available in the framework. 
These tests confirm that \colibri{} reproduces the expected behaviour of all methods, while highlighting the advantages of Bayesian inference as a principled and fully probabilistic characterisation of uncertainties. 
In addition, the ability to exploit posterior samples directly for correlation studies further demonstrates the flexibility and power of the Bayesian approach within the \colibri{} framework.

\colibri{} is intended as a living project. 
The code is under active development, with new features and inference strategies continuously being added. 
In addition to the core infrastructure, we also plan to provide ready-to-use PDF models as part of the distribution. 
One such example is already available from our recent study~\cite{Costantini:2025wxp}, 
which provides an implementation of a model that performs dimensional reduction of a neural network into its dominant modes, 
available at \url{https://github.com/HEP-PBSP/wmin-model}. 
Other models that are being developed are models based on Gaussian Processes~\cite{Candido:2024hjt}, models based on more 
realistic polynomial parametrisations and models based on neural networks. 
Users are encouraged to implement more models, e.~g. those based on flexible polynomial basis~\cite{Courtoy:2022ocu,Kotz:2025une} or advanced ML tools such as Bayesian neural networks.  
The online documentation is continuously updated and serves as the primary reference for recent progress, available functionalities, and tutorials.

In the near future, \colibri{} will be extended to fit PDFs simultaneously with SM precision parameters as well as 
with the Wilson Coefficients that parametrise New Physics degrees of freedom in some Effective Field Theories. 
Such a step is crucial, as the correlation between PDFs and SM precision parameters can no longer be ignored~\cite{Forte:2020pyp,Cooper-Sarkar:2020twv,Cridge:2023ztj,Alekhin:2024bhs,Cridge:2025wwo,Ball:2025xgq}, 
and the availability of open-source tools for 
simultaneous fits of PDFs and SM parameters would be a key advancement in keeping correlations into account.
In particular, while precise LHC data significantly enhance PDF precision, it has been shown in several recent publications~\cite{Carrazza:2019sec,Greljo:2021kvv,Madigan:2021uho,Gao:2022srd,Kassabov:2023hbm,Hammou:2023heg,Costantini:2024xae,Hammou:2024xuj} 
that they can also be sensitive to BSM dynamics. 
If BSM signals distort an experimental distribution included in a PDF fit, which is typically assumed
to follow the SM, this can lead to inconsistencies, which in turn can have PDFs adopting and incorporating the BSM effects, resulting in BSM-biased PDFs. 
A flexible tool for a simultaneous fit of PDFs and Wilson coefficients, in the same spirit but more general as compared to those provided by {\tt SimuNET}~\cite{Costantini:2024xae} or 
{\tt xFitter}~\cite{Shen:2024sci}, would avoid such bias and ensure that BSM effects and PDF effects can be disentangled.  
The extension of \colibri{} in both directions would provide the flexible tool that is needed to explore this completely new territory with 
a Bayesian inference method that allows full control over the role of prior assumptions.

The \colibri{} code is publicly available from its {\sc\small GitHub} repository:
\begin{center}
  {\tt \url{https://github.com/HEP-PBSP/colibri}},
\end{center}
and is accompanied by documentation and tutorials provided at:
\begin{center}
 \web{}.
\end{center}

\section*{Acknowledgements}
We thank Gaia Fontana (\verb|@qftoons|) for designing the \colibri{} logo.
We thank Ella Cole, Francesco Merlotti, Elie Hammou, Manuel Moreales Alvarado, David Yallup and the  
members of the NNPDF collaboration for insightful discussions. We are indebted to Juan Cruz Martinez for many his help with the NNPDF code, 
and Zahari Kassabov for his contributions during the earliest stages of the project.
Mark N.~Costantini and Maria Ubiali are supported
by the European Research Council under the European Union’s
Horizon 2020 research and innovation Programme (grant agreement
n.950246), and partially by the STFC
consolidated grant ST/X000664/1.
LM acknowledges support from the European Union under the MSCA fellowship (Grant agreement N. 101149078) {\it Advancing global SMEFT fits in the LHC precision era (EFT4ward)}. J. M. M. is supported by the donation of Christina and Peter Dawson to Lucy Cavendish College. Valentina Sch\"utze S\'{a}nchez is supported by the Newnham Scholarship for Women in Theoretical Physics. 
\newpage

\clearpage
\appendix

\addtocontents{toc}{%
  \protect\setlength{\protect\cftsecnumwidth}{7.0em}%
  \protect\setlength{\protect\cftbeforesecskip}{2pt}%
}
\section{Bayesian Update}
\label{app:bayesian_update}
Suppose experimental data comprising \(N_{\rm data}\) datapoints is distributed according to a multivariate normal:
\[
  \mathbf{D} \sim \mathcal{N}\bigl(\mathrm{FK}(\boldsymbol{\theta}),\,\Sigma\bigr),
\]
where \(\Sigma\) is the covariance matrix of dimension \(N_{\rm data}\times N_{\rm data}\).

In Bayesian statistics, \(\boldsymbol{\theta}\) itself is treated as a random variable with prior density \(\pi(\boldsymbol{\theta})\), here taken to be a sufficiently wide uniform distribution.  After observing \(\mathbf{D}_0\), Bayes' theorem yields the posterior
\begin{equation}
  \label{eq:bayes-theorem}
  p(\boldsymbol{\theta}\mid\mathbf{D}_0)
  = \dfrac{\pi(\boldsymbol{\theta})\,L(\mathbf{D}_0\mid\boldsymbol{\theta})}{Z}
  = \dfrac{\pi(\boldsymbol{\theta})
          \exp\!\left(-\tfrac{1}{2}\,\|\mathbf{D}_0-\mathrm{FK}(\boldsymbol{\theta})\|^2_{\Sigma}\right)}
         {\int d\boldsymbol{\theta}\;\pi(\boldsymbol{\theta})
          \exp\!\left(-\tfrac{1}{2}\,\|\mathbf{D}_0-\mathrm{FK}(\boldsymbol{\theta})\|^2_{\Sigma}\right)}\,.
\end{equation}
where we define the generalised \(L_2\) norm
\[
  \|\mathbf{x}\|^2_{\Sigma} = \mathbf{x}^T\,\Sigma^{-1}\,\mathbf{x},
  \quad\mathbf{x}\in\mathbb{R}^{N_{\rm data}}\,. 
\]

Now assume \(\mathbf{D}_0=(\mathbf{D}_1,\mathbf{D}_2)^T\) with 
\(\mathbf{D}_1\in\mathbb{R}^{n_1}\), \(\mathbf{D}_2\in\mathbb{R}^{n_2}\), \(n_1+n_2=N_{\rm data}\),
and that the two subsets are uncorrelated so that
\[
  \Sigma = \Sigma_1 \oplus \Sigma_2,
  \quad
  \Sigma_1\in\mathbb{R}^{n_1\times n_1},\;
  \Sigma_2\in\mathbb{R}^{n_2\times n_2}.
\]
The likelihood then factorises, and from \eqref{eq:bayes-theorem} we obtain
\begin{equation}
  \label{eq:bayes-uncorr}
  p(\boldsymbol{\theta}\mid\mathbf{D}_0)
  = \dfrac{\displaystyle
      \pi(\boldsymbol{\theta})
      \exp\!\bigl(-\tfrac12\|\mathbf{D}_1-\mathrm{FK}_1(\boldsymbol{\theta})\|^2_{\Sigma_1}\bigr)
      \exp\!\bigl(-\tfrac12\|\mathbf{D}_2-\mathrm{FK}_2(\boldsymbol{\theta})\|^2_{\Sigma_2}\bigr)
    }{\displaystyle
      \int d\boldsymbol{\theta}\;\pi(\boldsymbol{\theta})
      \exp\!\bigl(-\tfrac12\|\mathbf{D}_1-\mathrm{FK}_1(\boldsymbol{\theta})\|^2_{\Sigma_1}\bigr)
      \exp\!\bigl(-\tfrac12\|\mathbf{D}_2-\mathrm{FK}_2(\boldsymbol{\theta})\|^2_{\Sigma_2}\bigr)
    }\,,
\end{equation}
where \(\mathrm{FK}(\boldsymbol{\theta})=(\mathrm{FK}_1(\boldsymbol{\theta}),\mathrm{FK}_2(\boldsymbol{\theta}))^T\).

Next, note that the posterior given only \(\mathbf{D}_1\) is
\begin{equation}
  \label{eq:pd1-post}
  p_{\mathbf{D}_1}(\boldsymbol{\theta}\mid\mathbf{D}_1)
  = \dfrac{\pi(\boldsymbol{\theta})
          \exp\!\bigl(-\tfrac12\|\mathbf{D}_1-\mathrm{FK}_1(\boldsymbol{\theta})\|^2_{\Sigma_1}\bigr)}
         {\displaystyle
          \int d\boldsymbol{\theta}\;\pi(\boldsymbol{\theta})
          \exp\!\bigl(-\tfrac12\|\mathbf{D}_1-\mathrm{FK}_1(\boldsymbol{\theta})\|^2_{\Sigma_1}\bigr)
         }
  = \dfrac{\pi(\boldsymbol{\theta})
           \exp\!\bigl(-\tfrac12\|\mathbf{D}_1-\mathrm{FK}_1(\boldsymbol{\theta})\|^2_{\Sigma_1}\bigr)}
         {Z_1}\,.
\end{equation}
Substituting into \eqref{eq:bayes-uncorr} gives
\[
  p(\boldsymbol{\theta}\mid\mathbf{D}_0)
  = \dfrac{p_{\mathbf{D}_1}(\boldsymbol{\theta}\mid\mathbf{D}_1)\,
          \exp\!\bigl(-\tfrac12\|\mathbf{D}_2-\mathrm{FK}_2(\boldsymbol{\theta})\|^2_{\Sigma_2}\bigr)}
         {\displaystyle
          \int d\boldsymbol{\theta}\;p_{\mathbf{D}_1}(\boldsymbol{\theta}\mid\mathbf{D}_1)\,
          \exp\!\bigl(-\tfrac12\|\mathbf{D}_2-\mathrm{FK}_2(\boldsymbol{\theta})\|^2_{\Sigma_2}\bigr)
         }\,.
\]

If \(\mathbf{D}\sim\mathcal{N}(\mathrm{FK}(\boldsymbol{\theta}),\Sigma)\) with
\(\Sigma=\bigoplus_{i=1}^n\Sigma_i\), this argument applies recursively, yielding
\[
  p(\boldsymbol{\theta}\mid\mathbf{D}_0)
  = \dfrac{\displaystyle
      \prod_{i=1}^{n-1} p_{\mathbf{D}_i}(\boldsymbol{\theta}\mid\mathbf{D}_i)\,
      \exp\!\bigl(-\tfrac12\|\mathbf{D}_n-\mathrm{FK}_n(\boldsymbol{\theta})\|^2_{\Sigma_n}\bigr)
    }{\displaystyle
      \int d\boldsymbol{\theta}\;
      \prod_{i=1}^{n-1} p_{\mathbf{D}_i}(\boldsymbol{\theta}\mid\mathbf{D}_i)\,
      \exp\!\bigl(-\tfrac12\|\mathbf{D}_n-\mathrm{FK}_n(\boldsymbol{\theta})\|^2_{\Sigma_n}\bigr)
    }\,,
\]
with
\[
  p_{\mathbf{D}_k}(\boldsymbol{\theta}\mid\mathbf{D}_k)
  = \dfrac{\displaystyle
      \prod_{i=1}^{k-1} p_{\mathbf{D}_i}(\boldsymbol{\theta}\mid\mathbf{D}_i)\,
      \exp\!\bigl(-\tfrac12\|\mathbf{D}_k-\mathrm{FK}_k(\boldsymbol{\theta})\|^2_{\Sigma_k}\bigr)
    }{\displaystyle
      \int d\boldsymbol{\theta}\;
      \prod_{i=1}^{k-1} p_{\mathbf{D}_i}(\boldsymbol{\theta}\mid\mathbf{D}_i)\,
      \exp\!\bigl(-\tfrac12\|\mathbf{D}_k-\mathrm{FK}_k(\boldsymbol{\theta})\|^2_{\Sigma_k}\bigr)
    }.
\]

\addtocontents{toc}{%
  \protect\setlength{\protect\cftsecnumwidth}{7.0em}%
  \protect\setlength{\protect\cftbeforesecskip}{2pt}%
}
\section{Bayesian Linear Regression}
\label{app:bayes_lin_reg}
To illustrate the analytic method, let us assume a likelihood of the kind
\begin{equation}
  \label{eq:likelihood-general}
  p(D \mid \theta)
  = \dfrac{1}{(2\pi)^{N_{\rm dat}/2}\,\lvert\Sigma\rvert^{1/2}}
    \exp\!\Bigl(-\tfrac12\,(D - f(\theta))^T\,\Sigma^{-1}\,(D - f(\theta))\Bigr)\,,
\end{equation}
with $\theta$ being the parameters of the model, $D$ the data vector and $\Sigma$ its covariance. In general, a linear model can be described by the following equation:
\begin{equation}
  \label{eq:linear-model}
  f(\theta) = W\,\theta,
\end{equation}
where $W$ is a matrix that maps $\theta$ to the theory prediction vector, $f(\theta)$. For such a model, the likelihood factorises as:
\begin{align}
  \label{eq:likelihood-factorised}
  p(D \mid \theta)
  &= \dfrac{(2\pi)^{N_{\theta}/2}\,\bigl\lvert(W^T\Sigma^{-1}W)^{-1}\bigr\rvert^{1/2}}
         {(2\pi)^{N_{\rm dat}/2}\,\lvert\Sigma\rvert^{1/2}}
     \exp\!\Bigl(-\tfrac12\,(D - \hat{D})^T\,\Sigma^{-1}\,(D - \hat{D})\Bigr)\nonumber\\
  &\quad\times
     \dfrac{\exp\!\bigl(-\tfrac12\,(\theta - \hat{\theta})^T\,W^T\Sigma^{-1}W\,(\theta - \hat{\theta})\bigr)}
          {(2\pi)^{N_{\theta}/2}\,\bigl\lvert(W^T\Sigma^{-1}W)^{-1}\bigr\rvert^{1/2}}\nonumber\\
  &= (2\pi)^{N_{\theta}/2}\,\bigl\lvert(W^T\Sigma^{-1}W)^{-1}\bigr\rvert^{1/2}\;
     p(D \mid \hat{\theta})\,p(\hat{\theta}\mid\theta),
\end{align}
where $N_{\theta}$ is the number of parameters in the model, and we have defined the following relations;
\begin{equation}
  \label{eq:mle}
  \hat{\theta}
  = (W^T\,\Sigma^{-1}\,W)^{-1}\,W^T\,\Sigma^{-1}\,D,
  \quad
  \hat{D} = W\,\hat{\theta},
\end{equation}
and
\begin{equation}
  \label{eq:posterior-conditional}
  p(\hat{\theta}\mid\theta)
  = \dfrac{\exp\!\bigl(-\tfrac12\,(\theta - \hat{\theta})^T\,W^T\Sigma^{-1}W\,(\theta - \hat{\theta})\bigr)}
         {(2\pi)^{N_{\theta}/2}\,\bigl\lvert(W^T\Sigma^{-1}W)^{-1}\bigr\rvert^{1/2}}.
\end{equation}

With a uniform prior,
\begin{equation}
  \label{eq:uniform-prior}
  p(\theta_i)
  = \begin{cases}
      \displaystyle \frac{1}{b_i - a_i}, & \theta_i \in [a_i, b_i],\\[1.5ex]
      0,                    & \text{otherwise},
    \end{cases}
\end{equation}
the posterior becomes
\begin{align}
  \label{eq:posterior-uniform}
  p(\theta \mid D)
  &\propto p(D \mid \theta)\,p(\theta)\nonumber\\
  &\propto p(D \mid \theta)\,
     \prod_{i=1}^{N_{\theta}}
     \dfrac{\Theta(\theta_i - a_i)\,\Theta(b_i - \theta_i)}{b_i - a_i}.
\end{align}

\section{How to implement a PDF model in \colibri{}}
\label{app:model}

This appendix presents an example of how to implement a PDF model in \colibri{}. We will first show how to download the code, and then present an example of a model implementation, discussing the building blocks of a \colibri{} model. We will then show how to implement the Les Houches parametrisation as an example.

The content presented in this appendix is discussed in further detail in the \colibri{} documentation, \web{}.

\subsection{Installing \colibri{} on Linux or macOS}
\label{installation}

This section covers installing \colibri{} in various ways.

\subsubsection{Development Installation via Conda:}
You can install colibri easily by first cloning the repository and then using the provided \texttt{environment.yml} file:

\begin{lstlisting}[basicstyle=\ttfamily, xleftmargin=2em]
git clone https://github.com/HEP-PBSP/colibri
cd colibri
\end{lstlisting}
From your conda base environment run:

\begin{lstlisting}[basicstyle=\ttfamily, xleftmargin=2em]
conda env create -f environment.yml
\end{lstlisting}
This will create a \texttt{colibri-dev} environment installed in development mode.  
If you want to use a different environment name you can run:

\begin{lstlisting}[basicstyle=\ttfamily, xleftmargin=2em]
conda env create -n myenv -f environment.yml
\end{lstlisting}

\subsubsection{Installing with pip:}
If you don't want to clone the repository and don't need to work in development mode you can follow the installation instructions below.

Note that most of the \colibri{} dependencies are available in the \href{https://pypi.org/}{PyPi repository}. However non-python codes such as LHAPDF and pandoc won't be installed automatically and need to be manually installed in the environment. Because of this, we recommend the use of a conda environment. So the first step would be to create one from your base environment. For instance;

\begin{lstlisting}[basicstyle=\ttfamily, xleftmargin=2em]
conda create -n colibri-dev python>=3.11
\end{lstlisting}
In this new environment, install the following conda packages:

\begin{lstlisting}[basicstyle=\ttfamily, xleftmargin=2em]
conda install mpich lhapdf pandoc mpi4py ultranest pip
\end{lstlisting}
After having completed this, you can simply install the rest of the dependencies with \texttt{pip}:

\begin{lstlisting}[basicstyle=\ttfamily, xleftmargin=2em]
python -m pip install git+https://github.com/HEP-PBSP/colibri.git
\end{lstlisting}
Note that this will install the latest development version. If you want to install a specific release, you can specify the version. For instance, for v1.0.0, you can use the following command:

\begin{lstlisting}[basicstyle=\ttfamily, xleftmargin=2em]
python -m pip install git+https://github.com/HEP-PBSP/colibri.git@v1.0.0
\end{lstlisting}
To verify that the installation went through:

\begin{lstlisting}[basicstyle=\ttfamily, xleftmargin=2em]
python -c "import colibri; print(colibri.__version__)"
colibri --help
\end{lstlisting}

\subsubsection{GPU (CUDA) JAX Support}

The installation instructions shown above will install JAX in cpu mode. It is however possible to run \colibri{} fits using GPU cuda support too.  
To do so, after installing the package following one of the methods shown above, if you are on a Linux machine you can install JAX in CUDA mode by running:

\begin{lstlisting}[basicstyle=\ttfamily, xleftmargin=2em]
pip install -U "jax[cuda12]" -f https://storage.googleapis.com/jax-releases/jax_releases.html
\end{lstlisting}
(Note that this is a single command line). 

It is possible to run fits using float32 precision. The only way of doing so currently is to apply a patch to UltraNest so that the \texttt{json.dump} is compatible. To do that, run the following commands;

\begin{lstlisting}[basicstyle=\ttfamily, xleftmargin=2em]
git clone git@github.com:LucaMantani/UltraNest.git
cd UltraNest
git switch add-numpy-encoder
pip install .
\end{lstlisting}

\subsection{Implementing a model in \colibri{}}

In general, a \colibri{} model is contained in a directory with the following structure:

\begin{verbatim}

model_to_implement/
|-- pyproject.toml        # Defines a python package for the project and sets up executable
|-- model_to_implement/   
|   |-- app.py            # Enables the use of reportengine and validphys 
|   |-- config.py         # Defines the configuration layer for the model
|   |-- model.py          # Script where the model is defined
|-- runcards/             # Directory containing any runcards

\end{verbatim}

The best way to understand how to implement a model is to go through an example, so let's have a  look at how the Les Houches parametrisation is built. 

\subsection{Example: Les Houches parametrisation model}

In this section, we discuss how to implement a model in \colibri{}, using the Les Houches parametrisation model as an example. This parametrisation is simple enough for us to exemplify the use of \colibri{}, while still being realistic enough that this tutorial can be used as a template for other, more complex parametrisations or models.

Following this parametrisation, our basis has four PDFs, which in the evolution basis, read as in Eq.~\eqref{eq:evolution-basis-set}. 
After applying the Les Houches parametrisation assumptions and sum rules spelled out in~\ref{app:les-houches-param} we are left with 
13 free parameters to fit, namely $\alpha_g, \beta_g, \alpha_{u_v}, \beta_{u_v}, \epsilon_{u_v}, \gamma_{u_v}, \alpha_{d_v}, \beta_{d_v}, \epsilon_{d_v}, \gamma_{d_v}, \alpha_\Sigma, \beta_\Sigma$ and $A_g(A_\Sigma)$. We will now discuss how to implement this parametrisation in \colibri{}.

\subsubsection{Implementing the Les Houches model in Colibri}

In the \texttt{colibri/examples/} directory, you will find a directory called \texttt{les\_houches\_example}, which follows the structure defined above. We will have a look at them one by one. \\[0.3cm]
\hspace*{0.7cm}\underline{\textbf{\texttt{pyproject.toml}}}\mbox{}\\[0.3cm]
The \texttt{pyproject.toml} file defines the Python package configuration for this model using Poetry\cite{Eustace_Poetry_Python_packaging} as the dependency management and packaging tool. The configuration file structure looks like this:

\begin{lstlisting}[language=Python, xleftmargin=2em, caption={Example \texttt{pyproject.toml} script for the Les Houches parametrisation model.}, label={lst:pyproject}]

[build-system]
requires = ["poetry-core>=1.0.0", "poetry-dynamic-versioning>=1.1.0"]
build-backend = "poetry_dynamic_versioning.backend"

[tool.poetry]
name = "les_houches_example"
version = "1.0.0"
authors = ["PBSP collaboration"]
description = "Les Houches Parametrisation Example"

[tool.poetry.dependencies]


[tool.poetry.extras]
test = [
    "pytest",
    "hypothesis",
]
doc = [
    "sphinx",
    "recommonmark",
    "sphinx_rtd_theme"
]

[tool.poetry.scripts]
les_houches_exe = "les_houches_example.app:main"

\end{lstlisting}
Note that here the executable \texttt{les\_houches\_exe} is introduced, which is an executable that is specific to this model, and will be used to initialise a fit.\\[0.3cm]
\hspace*{0.7cm}\underline{\textbf{\texttt{app.py}}}\mbox{}\\[0.3cm]
The \texttt{app.py} module defines the core application class for the Les Houches model:

\begin{lstlisting}[language=Python, xleftmargin=2em, caption={Example \texttt{app.py} script for the Les Houches parametrisation model.}, label={lst:app.py}]

"""
les_houches_example.app.py

"""

from colibri.app import colibriApp
from les_houches_example.config import LesHouchesConfig


lh_pdf_providers = [
    "les_houches_example.model",
]


class LesHouchesApp(colibriApp):
    config_class = LesHouchesConfig


def main():
    a = LesHouchesApp(name="les_houches", providers=lh_pdf_providers)
    a.main()


if __name__ == "__main__":
    main()
    
\end{lstlisting}
The \texttt{LesHouchesApp} class enables the Les Houches model to function as a reportengine App\cite{zahari_kassabov_2019_2571601}. This integration provides a structured framework for data processing and report generation.

Some key features are:

\begin{itemize}
    \item \textbf{Provider System:} the \texttt{LesHouchesApp} accepts a list of providers (\texttt{lh\_pdf\_providers}) containing modules that are recognized by the application framework.
    \item \textbf{Inheritance Hierarchy:} the \texttt{LesHouchesApp} is a subclass of \texttt{colibriApp}, which means it automatically inherits all providers from both \colibri{} and validphys, giving access to their full functionality without the need for additional configuration.
\end{itemize}

\underline{\textbf{\texttt{config.py}}}\mbox{}\\[0.3cm]
The \texttt{config.py} script defines the configuration layer for the Les Houches model. It extends \colibri{}'s configuration system to provide a custom model builder and environment.

\begin{lstlisting}[language=Python, xleftmargin=2em, caption={Example \texttt{config.py} script for the Les Houches parametrisation model.}, label={lst:config.py}]

"""
les_houches_example.config.py

"""

import dill
import logging
from les_houches_example.model import LesHouchesPDF

from colibri.config import Environment, colibriConfig


log = logging.getLogger(__name__)


class LesHouchesEnvironment(Environment):
    pass


class LesHouchesConfig(colibriConfig):
    """
    LesHouchesConfig class Inherits from colibri.config.colibriConfig
    """

    def produce_pdf_model(self, output_path, dump_model=True):
        """
        Produce the Les Houches model.
        """
        model = LesHouchesPDF()
        # dump model to output_path using dill
        # this is mainly needed by scripts/bayesian_resampler.py
        if dump_model:
            with open(output_path / "pdf_model.pkl", "wb") as file:
                dill.dump(model, file)
        return model
        
\end{lstlisting}
The \texttt{produce\_pdf\_model} method creates an instance of the \texttt{LesHouchesPDF} model. Therefore, every model should have this production rule.

If \texttt{dump\_model} is set to \texttt{True}, the method serialises the model using \texttt{dill} and writes it to \texttt{pdf\_model.pkl} in the \texttt{output\_path}, where \texttt{output\_path} will be the output directory created when running a \colibri{} fit. \texttt{pdf\_model.pkl} will be loaded by \texttt{scripts/bayesian\_resampler.py} for resampling. 

If \texttt{dump\_model} is set to \texttt{False}, the serialised model will not be written to the disk.\\[0.3cm] 
\underline{\textbf{\texttt{model.py}}}\mbox{}\\[0.3cm]
The \texttt{model.py} script defines the Les Houches parametrisation model. It does so by defining the \texttt{LesHouchesPDF} class, which is based on the more abstract \texttt{PDFModel} class within \colibri{}. As described in Section \ref{sec:pdf-model-class}, it needs to specify a list of parameters to be fitted, and a map from these parameters to PDF values on a specified grid in $x$ for each flavour. In the case of the Les Houches parametrisation the parameter names are defined as follows:

\begin{lstlisting}[language=Python, xleftmargin=2em, caption={Example extract of a \texttt{model.py} script for the Les Houches parametrisation model, showing how \texttt{param\_names} is filled.}, label={lst:model.py-param-names}]

"""
les_houches_example.model.py
"""

import jax.numpy as jnp
import jax.scipy.special as jsp
from colibri.pdf_model import PDFModel


class LesHouchesPDF(PDFModel):
    """
    A PDFModel implementation for the Les Houches parametrisation.
    """

    @property
    def param_names(self):
        """The fitted parameters of the model."""
        return [
            "alpha_gluon",
            "beta_gluon",
            "alpha_up",
            "beta_up",
            "epsilon_up",
            "gamma_up",
            "alpha_down",
            "beta_down",
            "epsilon_down",
            "gamma_down",
            "norm_sigma",
            "alpha_sigma",
            "beta_sigma",
        ]

    @property
    def n_parameters(self):
        """The number of parameters of the model."""
        return len(self.param_names)

\end{lstlisting}

(Note that it would also be possible to automate the name generation for all parameters, for example by setting them all as \texttt{w\_i} for $i$ running over all parameters. This approach would be convenient for a parametrisation with many parameters.) 

For the map to PDF values, we define a function for each PDF, as described in Eq. \ref{eq:evolution-basis-set}. For example, for the gluon PDF, we have:

\begin{lstlisting}[language=Python, xleftmargin=2em, caption={Example extract of a \texttt{model.py} script for the Les Houches parametrisation model, showing how a map to PDF values is defined through functions that describe PDFs for each flavour.}, label={lst:model.py-pdf-functions}]

    def _pdf_gluon(
        self, x, alpha_gluon, beta_gluon, norm_sigma, alpha_sigma, beta_sigma
    ):
        """Computes normalisation factor A_g in terms of free parameters and computes the gluon PDF."""
        A_g = (
            jsp.gamma(alpha_gluon + beta_gluon + 2)
            / (jsp.gamma(alpha_gluon + 1) * jsp.gamma(beta_gluon + 1))
        ) * (
            1
            - norm_sigma
            * (jsp.gamma(alpha_sigma + 1) * jsp.gamma(beta_sigma + 1))
            / jsp.gamma(alpha_sigma + beta_sigma + 2)
        )
        return A_g * x**alpha_gluon * (1 - x) ** beta_gluon
        
\end{lstlisting}

The final building block of the \texttt{PDFModel} class is the \texttt{grid\_values\_func} function, which takes the parameters of the model and produces a grid that stores the PDF values for each point in $x$, where the values of $x$ are taken from the \texttt{xgrid}. Here we show how it is defined for the gluon PDF parameters, and skip the others for conciseness:

\begin{lstlisting}[language=Python, xleftmargin=2em]

    def grid_values_func(self, xgrid):
        """This function should produce a grid values function, which takes
        in the model parameters, and produces the PDF values on the grid xgrid.
        """

        xgrid = jnp.array(xgrid)

        def pdf_func(params):
            """ """
            alpha_gluon = params[0]
            beta_gluon = params[1]
            [...]       # other parameters
            pdf_grid = []

            # Compute the PDFs for each flavour
            gluon_pdf = self._pdf_gluon(
                xgrid, alpha_gluon, beta_gluon, norm_sigma, alpha_sigma, beta_sigma
            )

            [...]       # sigma_pdf, valence3_pdf, t8_pdf
            
            # Build the PDF grid
            pdf_grid = jnp.array(
                [
                    jnp.zeros_like(xgrid),  # Photon
                    sigma_pdf,  # \Sigma
                    gluon_pdf,  # g
                    valence_pdf,  # V
                    valence3_pdf,  # V3
                    valence_pdf,  # V8 = V
                    valence_pdf,  # V15 = V
                    valence_pdf,  # V24 = V
                    valence_pdf,  # V35 = V
                    jnp.zeros_like(xgrid),  # T3 = 0
                    t8_pdf,  # T8
                    sigma_pdf,  # T15 = \Sigma
                    sigma_pdf,  # T24 = \Sigma
                    sigma_pdf,  # T35 = \Sigma
                ]
            )
            return pdf_grid

        return pdf_func
        
\end{lstlisting}

The \texttt{LesHouchesPDF} class completes the abstract methods of the \texttt{PDFModel} class. This allows for the definition of a
specific model in a way that can be used in the \colibri{} code. The \texttt{LesHouchesPDF} class does the following:

\begin{itemize}
\item takes a list of flavours to be fitted (\texttt{param\_names}), 
\item defines the PDF for each flavour, 
\item  computes grid values. 
\end{itemize}

Having defined this model, it is used in the production rule \texttt{produce\_pdf\_model}, defined in the \texttt{config.py} script, shown above. This allows the model to be seen by the rest of the code, so that it can be used to run a fit and perform closure tests.\\[0.3cm]
\underline{\textbf{Installing a model and running a fit}}\\[0.3cm]
As mentioned above, each model should have its own \texttt{pyproject.toml} script, which defines the Python package configuration for this model. Each model can be installed by running 
\begin{lstlisting}[basicstyle=\ttfamily, xleftmargin=2em]
pip install -e .
\end{lstlisting}
in the model directory, which is the one where \texttt{pyproject.toml} should be.

This will set up a model-specific executable, which can be used to run fits. In the case of the Les Houches model, this executable is \texttt{les\_houches\_exe}, and can be run simply as:

\begin{lstlisting}[basicstyle=\ttfamily, xleftmargin=2em]
les_houches_exe my_runcard.yml
\end{lstlisting}

which will produce a directory called \texttt{my\_runcard} with the results. More information on runcard settings and how to process results can be found in the \colibri{} documentation.

\section{The Les Houches Parametrisation}
\label{app:les-houches-param}

\subsection{Free parameters in the Les Houches Parametrisation}
We adopt the Les Houches parametrisation taken from ref.~\cite{Alekhin:2005xgg}, where it is assumed that the total sea, $\Sigma=u+\bar{u}+d+\bar{d}+s+\bar{s}$, is constrained to be made 40\% by up and anti-up, 40\% by down and anti-down, and 20\% by strange and anti-strange, which means that we can write:
\begin{align}
    u+\bar{u}=0.4\Sigma,\notag\\
    d+\bar{d}=0.4\Sigma,\\
    s+\bar{s}=0.2\Sigma.\notag
\end{align}
It is also assumed that there is no difference between $\bar{u}$ and $\bar{d}$, so we are only left with four active flavours, namely $g, u_{v}, d_{v}$ and $\Sigma$. Furthermore, $\epsilon_g$, $\gamma_g$, $\epsilon_\Sigma$ and $\gamma_\Sigma$ are all set to zero. We are therefore left with the set of equations:
\begin{align}
\label{eq:flavour-basis-set}
xf_g(x,Q_0) &= A_g\,x^{\alpha_g}\,(1 - x)^{\beta_g} \notag \\
xf_{u_v}(x,Q_0) &=  A_{u_v}\,x^{\alpha_{u_v}}\,(1 - x)^{\beta_{u_v}}\, (1+\epsilon_{u_v}\sqrt{x}+\gamma_{u_v} x) \\
xf_{d_v}(x,Q_0) &= A_{d_v}\,x^{\alpha_{d_v}}\,(1 - x)^{\beta_{d_v}}\, (1+\epsilon_{d_v}\sqrt{x}+\gamma_{d_v} x)  \notag \\
xf_\Sigma(x,Q_0) &=  A_\Sigma\,x^{\alpha_\Sigma}\,(1 - x)^{\beta_\Sigma}.\notag
\end{align}
This amounts to 16 parameters. Moreover not all parameters are independent. $A_g$ is related to $A_\Sigma$ by the momentum sum rules:
\begin{equation}
\label{eq:gluon-sum-rule}
A_g\int_0^1 x^{\alpha_g}\,(1 - x)^{\beta_g} dx + A_\Sigma \int_0^1 x^{\alpha_\Sigma}\,(1 - x)^{\beta_\Sigma}\, dx = 1,
\end{equation}
and the $A_{u_v}$ and $A_{d_v}$ parameters are determined by the valence sum rules:
\begin{align}
\label{eq:valence-sum-rules}
& A_{u_v}\,\int\, x^{\alpha_{u_v}-1}\,(1 - x)^{\beta_{u_v}}\, (1+\epsilon_{u_v}\sqrt{x}+\gamma_{u_v} x)  dx = 2 \notag \\
& A_{d_v}\,\int\,x^{\alpha_{d_v}-1}\,(1 - x)^{\beta_{d_v}}\, (1+\epsilon_{d_v}\sqrt{x}+\gamma_{d_v} x) dx = 1,
\end{align}
leaving 13 free parameters\footnote{In ref.~\cite{Alekhin:2005xgg}, $\epsilon_{u_v}$ is fixed to its best-fit value, $\epsilon_{u_v} = -1.56$, in order to avoid instability due to a very high correlation between $u_v$ parameters. They therefore left only 12 parameters free to vary. We decide to leave $\epsilon_{u_v}$ free because we don't believe we will encounter this problem. }.

\subsection{Normalisations}
We can write the expressions for $A_g$, $A_{u_v}$ and $A_{d_v}$ explicitly by solving the integral spelled out in the sum rules, Equations \ref{eq:gluon-sum-rule} and \ref{eq:valence-sum-rules}, which are of the form of Euler beta functions, given by:

\begin{align*}
\label{eq:euler-beta-func}
\int_0^1 dt \, t^{v-1} (1-t)^{w -1} = \frac{\Gamma(v) \Gamma(w)}{\Gamma(v + w)},
\end{align*}
where, for positive integer $n$, $\Gamma(n)$ is defined as:
\begin{align*}
    \Gamma(n) = (n-1)!.
\end{align*}
We find that:
\begin{align*}
    A_g = \frac{\Gamma(\alpha_g + \beta_g + 2)}{\Gamma(\alpha_g+1)\Gamma(\beta_g+1)}\left[ 1 - A_{\Sigma} \frac{\Gamma(\alpha_\Sigma + 1) \Gamma(\beta_\Sigma + 1)}{\Gamma(\alpha_\Sigma + \beta_\Sigma +2)} \right],
\end{align*}
\begin{align*}
    A_{u_v} = \frac{2}{\Gamma(\beta_{u_v}+1)}{\left[ \frac{\Gamma(\alpha_{u_v})}{\Gamma(\alpha_{u_v} + \beta_{u_v} + 1)}  + \epsilon_{u_v} \frac{\Gamma(\alpha_{u_v} + 1 / 2)}{\Gamma(\alpha_{u_v} + \beta_{u_v} + 3 / 2)} + \gamma_{u_v} \frac{\Gamma(\alpha_{u_v} + 1)}{\Gamma(\alpha_{u_v} + \beta_{u_v} + 2)} \right]}^{-1},
\end{align*}
\begin{align*}
    A_{d_v} = \frac{1}{\Gamma(\beta_{d_v}+1)}{\left[ \frac{\Gamma(\alpha_{d_v})}{\Gamma(\alpha_{d_v} + \beta_{d_v} + 1)}  + \epsilon_{d_v} \frac{\Gamma(\alpha_{d_v} + 1 / 2)}{\Gamma(\alpha_{d_v} + \beta_{d_v} + 3 / 2)} + \gamma_{d_v} \frac{\Gamma(\alpha_{d_v} + 1)}{\Gamma(\alpha_{d_v} + \beta_{d_v} + 2)} \right]}^{-1},
\end{align*}

\subsection{The Les Houches Parametrisation in the evolution basis}
We can then use these expressions to find the elements of the evolution basis explicitly, which is given by: 

\begin{align}
\Sigma &= u+\bar{u}+d+\bar{d}+s+\bar{s}, \notag  \\
T_3 &= (u + \bar{u}) - (d + \bar{d}), \notag \\
T_8 &= (u+\bar{u} + d + \bar{d}) - 2(s+\bar{s}),\\
V &= (u-\bar{u}) + (d-\bar{d}) + (s-\bar{s}), \notag\\
V_3 &= (u - \bar{u}) - (d - \bar{d}), \notag\\
V_8 &= (u-\bar{u} + d - \bar{d}) - 2(s-\bar{s}). \notag
\end{align}

Noting that $u_v = u - \bar{u}$, $d_v = d - \bar{d}$ and that, since there are no valence strange quarks, $s_v = s - \bar{s} = 0$, and applying the assumptions stated above, we find:

\begin{align}
    T_3 &= 0.4\Sigma - 0.4\Sigma = 0, \notag \\
    T_8 &= 0.4\Sigma + 0.4\Sigma - 2 \cdot (0.2\Sigma) = 0.4\Sigma, \notag \\
    T_{15} &= T_{24} = T_{35} = \Sigma, \\
    V_8 &= u_v + d_v - 2 \cdot 0 = V, \notag \\
    V_{15} &= V_{24} = V_{35} = V. \notag
\end{align}
Therefore, we are again left with only four active flavours; $\Sigma, V, V_3$ and the gluon.
We already have an explicit parametrisation for $f_\Sigma$ and $f_g$, as stated in Eq. \ref{eq:flavour-basis-set}. We have the ingredients to write analogous expressions for $f_V$ and $f_{V_3}$, which are given by:
\begin{align}
    x f_V &= x f_{u_v} + x f_{d_v} \\
    &= A_{u_v}\,x^{\alpha_{u_v}}\,(1 - x)^{\beta_{u_v}}\, (1+\epsilon_{u_v}\sqrt{x}+\gamma_{u_v} x) + A_{d_v}\,x^{\alpha_{d_v}}\,(1 - x)^{\beta_{d_v}}(1+\epsilon_{d_v}\sqrt{x}+\gamma_{d_v} x)\notag  
\end{align}
\begin{align}
    x f_{V_3} &= x f_{u_v} - x f_{d_v} \\
    &= A_{u_v}\,x^{\alpha_{u_v}}\,(1 - x)^{\beta_{u_v}}\, (1+\epsilon_{u_v}\sqrt{x}+\gamma_{u_v} x) - A_{d_v}\,x^{\alpha_{d_v}}\,(1 - x)^{\beta_{d_v}}(1+\epsilon_{d_v}\sqrt{x}+\gamma_{d_v} x)\notag
\end{align}

\bibliographystyle{spphys}       


\end{document}